\documentclass[3p,onecolumn]{elsarticle}

\usepackage{hyperref}
\usepackage[utf8]{inputenc}
\usepackage[T1]{fontenc}
\usepackage{csquotes}
\usepackage[english]{babel}
\usepackage{graphicx}
\usepackage{adjustbox}
\usepackage{textcomp}
\usepackage{xcolor}
\usepackage{xspace}
\usepackage{tcolorbox}
\usepackage{booktabs}
\usepackage{arydshln}
\usepackage{siunitx}
\usepackage{balance}
\usepackage{url}
\usepackage{tikz}
\usepackage{fontawesome}
\usepackage{todonotes}
\usepackage{cleveref}
\usepackage{listings}
\usepackage{adjustbox}
\usepackage{color, colortbl}
\usepackage{tikz}
\usepackage{caption}
\usepackage{subcaption}
\usepackage{natbib}
\setcitestyle{sort&compress}
\usepackage[normalem]{ulem}
\useunder{\uline}{\ul}{}
\usepackage[resetlabels,labeled]{multibib}
\newcites{R}{resources}

\let\oriCiteO\citeR

\RenewDocumentCommand{\citeR}{O{} O{} m}{%
  \renewcommand{\citenumfont}[1]{R##1}%  
  \oriCiteO[#1][#2]{#3}%
  \renewcommand{\citenumfont}[1]{##1}%
}

\journal{Information and Software Technology}

\newcommand{\ie}{i.e.,\xspace}
\newcommand{\eg}{e.g.,\xspace}

\newcommand{\etal}{\emph{et al.}\xspace}

\usepackage{enumitem}

\definecolor{gray50}{gray}{.5}
\definecolor{gray40}{gray}{.6}
\definecolor{gray30}{gray}{.7}
\definecolor{gray20}{gray}{.8}
\definecolor{gray10}{gray}{.9}
\definecolor{gray05}{gray}{.95}

\newlength\Linewidth
\def\findlength{\setlength\Linewidth\linewidth
	\addtolength\Linewidth{-4\fboxrule}
	\addtolength\Linewidth{-3\fboxsep}
}

\definecolor{pblue}{rgb}{0.13,0.13,1}
\definecolor{pgreen}{rgb}{0,0.5,0}
\definecolor{pred}{rgb}{0.9,0,0}
\definecolor{pgrey}{rgb}{0.46,0.45,0.48}
\definecolor{codebackground}{rgb}{0.95, 0.95, 0.92}
\definecolor{arsenic}{rgb}{0.23, 0.27, 0.29}
\definecolor{mintgreen}{HTML}{cce3de}

\newtcolorbox{summarybox}[1][RQ]{
	colback=gray!5,
	colframe=arsenic,
	boxrule=0.4mm,
	left=1.5mm,
	right=1.5mm,
	top=1.5mm,
	bottom=1.5mm,
	fonttitle=\bfseries,
	title=Main findings for #1
}

\newcommand{\nb}[2]{
	\fbox{\bfseries\sffamily\scriptsize\color{red}#1}
	{\sf\small$\blacktriangleright${\color{red}\textit{#2}}$\blacktriangleleft$}
}

\renewcommand{\arraystretch}{1.2}

\definecolor{aliceblue}{rgb}{0.94, 0.97, 1.0}

\RequirePackage{fontawesome}
\newcommand{\keyfindingsone}[1]{ %
	\vspace{5pt} %
	\noindent\fcolorbox{arsenic}{mintgreen}{%
		\parbox{0.97\linewidth}{% 
			\textbf{\faKey\ Take Away Message.} #1 %
		}%
	}%
	\vspace{5pt} %
}%

\newcommand{\qsharp}{\textsc{Q$\sharp$}\xspace}

\definecolor{codegreen}{rgb}{0,0.6,0}
\definecolor{codegray}{rgb}{0.5,0.5,0.5}
\definecolor{codepurple}{rgb}{0.58,0,0.82}
\definecolor{backcolour}{rgb}{0.95,0.95,0.92}
\definecolor{Gray}{gray}{0.93}

\lstdefinestyle{mystyle}{
	backgroundcolor=\color{backcolour},   
	commentstyle=\color{codegreen},
	keywordstyle=\color{magenta},
	numberstyle=\tiny\color{codegray},
	stringstyle=\color{codepurple},
	basicstyle=\ttfamily\footnotesize,
	breakatwhitespace=false,         
	breaklines=true,                 
	captionpos=b,                    
	keepspaces=true,                 
	numbers=left,                    
	numbersep=5pt,                  
	showspaces=false,                
	showstringspaces=false,
	showtabs=false,                  
	tabsize=2
}

\newcommand{\rb}[1]{
	
	\begin{tcolorbox}[colback=gray!05,%gray background
		colframe=black,% black frame colour
		width=\columnwidth,% Use 5cm total width,
		arc=3mm, auto outer arc,
		boxrule=0.5pt,
		]
		#1
	\end{tcolorbox}
}

\newcounter{Finding}
\stepcounter{Finding}

\newcommand{\resq}[1]{
    \textbf{RQ$_{#1}$}
}

\newcommand{\resquestion}[2]{ %
    \vspace{5pt} %
    \noindent\fcolorbox{black}{blue!03}{%
        \parbox{0.97\linewidth}{% 
            \faSearch #1 %
            #2
        }%
    }%
    \vspace{5pt} %
}%

\newcommand{\slrguidelines}{\cite{keele2007guidelines, petersen2008systematic, petersen2015guidelines}}
\renewcommand{\arraystretch}{1.5}
\begin{document}

\begin{frontmatter}

\title{The Quantum Frontier of Software Engineering:\\A Systematic Mapping Study}

\author[mymainaddress]{Manuel De Stefano}
\cortext[mycorrespondingauthor]{Corresponding author}
\ead{madestefano@unisa.it}

\author[secondaddress]{Fabiano Pecorelli}
\ead{f.pecorelli@jads.nl}

\author[mymainaddress]{Dario Di Nucci}
\ead{ddinucci@unisa.it}

\author[mymainaddress]{Fabio Palomba}
\ead{fpalomba@unisa.it}

\author[mymainaddress]{Andrea De Lucia}
\ead{adelucia@unisa.it}

\address[mymainaddress]{SeSa Lab - University of Salerno, Italy}
\address[secondaddress]{Jheronimus Academy of Data Science - Eindhoven University of Technology, The Netherlands}

\begin{abstract}
	\textbf{Context.} 
Quantum computing is becoming a reality, and quantum software engineering (QSE) is emerging as a new discipline to enable developers to design and develop quantum programs. 
\textbf{Objective.}
This paper presents a systematic mapping study of the current state of QSE research, aiming to identify the most investigated topics, the types and number of studies, the main reported results, and the most studied quantum computing tools/frameworks.
Additionally, the study aims to explore the research community's interest in QSE, how it has evolved, and any prior contributions to the discipline before its formal introduction through the Talavera Manifesto.
\textbf{Method.}
We searched for relevant articles in several databases and applied inclusion and exclusion criteria to select the most relevant studies.
After evaluating the quality of the selected resources, we extracted relevant data from the primary studies and analyzed them.
\textbf{Results.}
We found that QSE research has primarily focused on software testing, with little attention given to other topics, such as software engineering management.
The most commonly studied technology for techniques and tools is Qiskit, although, in most studies, either multiple or none specific technologies were employed. 
The researchers most interested in QSE are interconnected through direct collaborations, and several strong collaboration clusters have been identified. 
Most articles in QSE have been published in non-thematic venues, with a preference for conferences.
\textbf{Conclusions.}
The study's implications are providing a centralized source of information for researchers and practitioners in the field, facilitating knowledge transfer, and contributing to the advancement and growth of QSE.
\end{abstract}


\end{frontmatter}

\section{Introduction}
Quantum computing will likely become the next concrete asset for researchers and practitioners. 
Every developer can now access a quantum computer and use quantum computation to solve arbitrary, computationally-intensive problems ~\cite{knight2018serious,hoare2005grand}.
This result is possible because of a great effort made by major software companies, like \textsc{IBM} and \textsc{Google}, which are currently investing hundreds of millions of dollars every year to produce novel hardware and software technologies that can support the execution of quantum programs, making a step further in democratizing quantum computing \cite{omer2003qcl,qsharp,altenkirch2005qml, aleksandrowicz2019qiskit,broughton2020tensorflow,steiger2018projectq}. 

Quantum programming is promising in resolving problems in various fields, such as machine learning, optimization, cryptography, and chemistry. However, the development of large-scale quantum software is still distant \cite{biamonte2017quantum, guerreschi2017practical, mailloux2016post, reiher2017elucidating}. 
To address this, researchers have proposed a new discipline, quantum software engineering (QSE), that extends classical software engineering into the quantum domain \cite{ piattini2021toward, piattini2020talavera, piattini2020quantum, moguel2020roadmap}.
This discipline aims to enable developers to design and develop quantum programs with the same confidence as classical programs, providing them with all the methods and tools necessary \cite{zhao2020quantum}.

Since the publication of the Talavera Manifesto \cite{piattini2020talavera}, which can be considered the pillar of quantum software engineering, many studies have been published, proposing novel approaches, tools, and techniques.
Consequently, several secondary studies have been published exploring various aspects of QSE, such as optimization \cite{yarkoni2022quantum, shi2020resource}, industrial adoption issues \cite{awan2022quantum}, testing \cite{garcia2021quantum}, and architecture \cite{ahmad2022towards}, to put order among all the published studies, and steer further research.
These previous studies, however, focused on specific aspects of the discipline, like testing \cite{garcia2021quantum}, and have failed to provide a comprehensive overview of the field.
As a result, significant research angles like the management and maintenance of quantum-based solutions have been left largely unexplored. 
To fully understand the potential of this field, and maximize the benefits obtainable from the current evidence \cite{kitchenham2004evidence}, a comprehensive synthesis of the latest research is essential.

In this paper, we propose a systematic mapping study on the current status of QSE research to fulfill this gap.
This systematic mapping study aims to provide a broad and holistic overview of QSE, determine its achievements, and identify current research gaps.
To this end, the study addresses several research questions about QSE research aspects.
In particular, our study aims to identify the most investigated topics in QSE, the types and the number of proposed studies, the main reported results, and the most studied quantum computing tools/frameworks. 
Additionally, the study explores the research community's interest in QSE, how it has evolved, and any prior contributions to the discipline before its formal introduction through the Talavera Manifesto \cite{piattini2020talavera}.
Moreover,  this study aims to understand the main researchers involved in the field, their research groups, their interactions, and their distribution concerning various SE topics.
The study also intends to identify which venues will most likely publish QSE articles outside thematic workshops.

The results of our systematic mapping study can provide valuable insights into the development and evolution of the research community, the tools and frameworks being studied, and the distribution of research topics among different software engineering areas. 
Our final goal is also to facilitate knowledge transfer and provide a centralized source of information for researchers and practitioners in the field, contributing to the advancement and growth of quantum software engineering.

The remainder of the paper is structured as follows: in \Cref{sec:related} related literature is discussed and analyzed; \Cref{sec:design} and \Cref{sec:results} present the research method employed to conduct our study and the achieved results, which are then discussed in \Cref{sec:discussion}; in \Cref{sec:conclusion}, then, final remarks and future research directions are given.

\section{Related Work} \label{sec:related}
Quantum Software Engineering (QSE) has recently attracted increasing attention.
Hence, several secondary studies were published, exploring different aspects of this discipline.
This section compares this literature with our secondary study, highlighting the differences and complementary points.

Yarkoni \etal \cite{yarkoni2022quantum} provided a literature review of quantum annealing (QA) technology and its applications. 
The study aims to provide a centralized source of information on the applications of QA and identify the advantages, limitations, and potential of the technology for both researchers and practitioners. 

Shi \etal \cite{shi2020resource} published a study that reviews quantum software optimization techniques, focusing on the efficiency of quantum computing (QC) systems. 
The study argues that greater efficiency of QC systems can be achieved by breaking the abstractions between layers in the quantum software stack. 

Awan \etal \cite{awan2022quantum} published a study that aims to identify and prioritize the most critical challenges facing the software industry in adopting quantum computing technology. 
The study implements a fuzzy analytic hierarchy process (F-AHP) to evaluate and rank the identified challenges. 
The results show that the critical barriers to quantum computing adoption are the lack of technical expertise, information accuracy, organizational interest in adopting the new process, and the lack of standards of secure communication. 

Garcìa de la Barrera \etal \cite{garcia2021quantum} conducted a systematic mapping study on quantum software testing.
The research provides a thorough overview of the current state of the art in quantum software testing.
It identifies two major trends in testing techniques: statistical approaches based on repeated measurements and the use of Hoare-like logic for reasoning about software correctness.
Reversible circuit testing for quantum software unitary testing is also mentioned by the authors, which is partially applicable to this domain.

Ahmad \etal \cite{ahmad2022towards} published a secondary study that addresses the challenges of architecture-centric implementation of quantum software systems. 
To achieve this, it explores the role of quantum software architectures (QSA) in supporting quantum software systems' design, development, and maintenance phases. 
The research focuses on two main aspects: (i) the architectural process with its architecting activities and (ii) the human roles that can leverage available tools to automate and customize the architecture-centric implementation of quantum software.
By examining these aspects, the study aims to provide insights into the principles of quantum mechanics and software engineering practices and how they can be integrated into the design and development of quantum software systems.

Reasoning on the current state of the art, we point out that previous systematic literature reviews and mapping studies have focused on specific themes, e.g., testing or architectures, rather than providing a comprehensive view of the research conducted on quantum software engineering. As a consequence, there is still little knowledge on the set of topics more frequently investigated by quantum software engineering researchers, other than of the main results achieved and how these can drive future research. The findings of our study can provide valuable insights into the development and evolution of the QSE research community, the tools and frameworks being studied, and the distribution of research topics among different software engineering areas. In addition, we also identify a gap of knowledge with respect to the venues and research groups that are currently investigating quantum software engineering aspects: these pieces of information are crucial for newcomers, e.g., fresh Ph.D. students, who would like to embrace the quantum software engineering field, other than more senior researchers interested in understanding the dynamics behind the quantum software engineering research community. 

%On the one hand, our mapping study is broader and holistic and does not limit the scope of the research to the sole software architecture field.
%On the other hand, our systematic mapping study on Quantum Software Engineering stands out as a comprehensive and holistic overview of the discipline, considering the entire field and not limiting the scope to specific aspects. 
%Our study is broader in scope compared to other secondary studies depicted above.
%The results of our study can provide valuable insights into the development and evolution of the QSE research community, the tools and frameworks being studied, and the distribution of research topics among different software engineering areas.
%Additionally, our study can facilitate knowledge transfer and provide a centralized source of information for researchers and practitioners in the field, contributing to the advancement and growth of QSE.
\section{Research Method} \label{sec:design}
The \emph{goal} of our study was to produce an organic and holistic view of the scientific literature in the field of quantum software engineering, with the \emph{purpose} of understanding what researchers' efforts have focused on, what needs further investigation, what has been achieved so far, and what the research gaps are. The \emph{perspective} is of both researchers and practitioners. The former are interested in having a unique source of information providing a comprehensive view of the current research in quantum software engineering. The latter are interested in understanding the current trends and technologies researchers produce that might be potentially transferred in practice. Our systematic mapping study has been conducted following the guidelines in \slrguidelines. In terms of reporting, we followed the \textsl{ACM/SIGSOFT Empirical Standards}\footnote{Available at \url{https://github.com/acmsigsoft/EmpiricalStandards}. Given the nature of our study and the currently available standards, we followed the \textsl{``General Standard''} and \textsl{``SystematicReviews''} definitions and guidelines.} 

\subsection{Research Questions}
We aimed to answer various research questions to achieve a holistic view of the state of quantum software engineering.
To understand and identify possible gaps in QSE research, especially in terms of coverage of various aspects of SE and the degree of maturity achieved in these areas, we asked:

\resquestion{ \resq{1} \textbf{Current research trends in QSE}}{
    \begin{description}
        \item \resq{1.1} \textit{How many and what kind of studies have been proposed in QSE?}
        \item \resq{1.2} \textit{Which areas of software engineering have received the most attention in QSE?}
        \item \resq{1.3} \textit{Which types of studies are most commonly proposed in the different areas of software engineering within QSE?}
    \end{description}
}

Answering RQ$_{1.1}$ will allow us to assess the degree of maturity of the research: a prevalence of philosophical papers depicts a different level of maturity than that of evaluation research papers.
The answer to RQ$_{1.2}$ will allow an understanding of possible gray areas or unexplored areas which should need further attention by the research community.
By combining the first two insights, answering RQ$_{1.3}$ will allow assessing the maturity of research production in each area and evaluating the degree of development of the specific area.

To understand the research progresses in QSE and identify possible research opportunities, we asked:

\resquestion{ \resq{2} \textbf{Achieved results and studied technologies}}{
    \begin{description}
        \item \resq{2.1}{What are the main results reported?}
        \item \resq{2.2}{Which quantum computing tools/frameworks are most being studied?}
    \end{description}
}

Answering the first question will allow us to understand the main results reported in the literature, supplying a picture of what is being studied, what has been discovered, and what is still unknown.
Answering the second question will provide information on the types of quantum computing tools/frameworks being studied to better understand the maturity of different implementations. 
This information can help us to understand the current state of quantum computing and its research landscape.

The Talavera Manifesto was introduced in 2020, yet quantum software has been discussed since before. The following research questions are posed to analyze the interest in this new discipline:

\resquestion{ \resq{3} \textbf{Evolution of QSE}}{
    \begin{description}
        \item \resq{3.1}{Are there contributions to QSE before the discipline was even born?}
        \item \resq{3.2}{How has the research community's interest in this new discipline evolved?}
    \end{description}
}
Answering these questions can provide valuable insights into the historical and evolutionary context of QSE, which can be beneficial in several ways.
First, understanding the contributions to QSE before the discipline was even born can help researchers appreciate the foundational ideas and concepts that underlie the field. 
Second, examining how the research community's interest in QSE has evolved over time can provide insights into the current state of the field and its trajectory. 
By understanding the factors that have driven the growth and development of QSE, researchers can identify areas that require further investment and attention to continue advancing the field.

To understand who the leading researchers involved in this discipline are and how they and their research groups interact, the following research questions are posed:

\resquestion{ \resq{4} \textbf{Authors and collaborations in QSE}}{
    \begin{description}
        \item \resq{4.1}{Who are the researchers most interested in QSE? How are they interconnected?}
        \item \resq{4.2}{Given the various SE topics, how are these researchers distributed?}
    \end{description}
}

Answering these questions can be valuable because it can provide a comprehensive understanding of the current landscape of QSE research and the key players in the field. 
Knowing who the leading researchers are can help identify the most influential and significant work in QSE. Understanding how these researchers and their groups are interconnected can provide insight into how ideas and research are shared and disseminated within the discipline. 
Understanding the relationships and interactions within the QSE research community can also help identify potential barriers or challenges to collaboration and ways to overcome them.

To better understand whether research in QSE is emerging from niche topics and to understand which venues are most attractive, the following research question is posed:

\resquestion{ \resq{5} \textbf{Publication trends in QSE}}{
    \begin{description}
        \item \resq{5.1}{Which venues will most likely publish QSE articles outside thematic venues?}
    \end{description}
}

Answering this question can be valuable because it can provide insight into the reach and impact of QSE research beyond specialized workshops and conferences. 
Knowing which venues are most likely to publish QSE articles can help identify the most influential and visible outlets for this type of research and inform strategies for publishing and disseminating work in QSE.
Understanding which venues are most attractive to QSE researchers can also help identify trends and patterns in the field and inform decisions about where to submit work for publication. 
In general, understanding the venues that are most likely to publish QSE articles can help to establish the reach and impact of this type of research and inform strategies for publishing and disseminating work in the field.
It must be noted that for \textit{tematic venues}, we intend venues explicitly focused on quantum software engineering or quantum software development.

% \usepackage{booktabs}
\begin{table}[t]
\caption{The number of studies per database.}
\label{tab:studies}
\centering
\begin{tabular}{@{}lr@{}}
\toprule
Database       & Search Results  \\ \midrule
ACM            & 1,376           \\
IEEE           & 1,991           \\
SCOPUS         & 998             \\
Web of Science & 1,002           \\\bottomrule
\end{tabular}
\end{table}

\subsection{Search Criteria}
Guidelines \slrguidelines suggest that only the variables of population and intervention are necessary to conduct our mapping study.
Comparison and outcome are unnecessary for a systematic mapping study because it is not a meta-analysis, randomized controlled trial, or systematic review \cite{petersen2008systematic, petersen2015guidelines}. 
Instead, it is a preliminary study that provides an overview of the current state of research in a particular area. 
A systematic mapping study aims to identify gaps in the existing literature and provide a general understanding of the topics, the methods, and the results in a specific field \cite{petersen2008systematic}. 
Therefore, the emphasis is on identifying and categorizing the existing literature rather than comparing or determining outcomes.

We developed the search strings utilizing these two variables (\ie population and intervention) as the primary keywords to narrow the scope of our research.
In software engineering, the term \textit{population} (P) may refer to various subgroups within the field, such as specific software engineering roles, categories of software engineers, application areas, or industry groups. 
These populations may have unique characteristics or challenges relevant to the research being conducted \cite{keele2007guidelines}.
In particular, the population in our context refers to papers focusing on quantum software engineering.
In the context of software engineering, the term \textit{intervention} (I) refers to any method, tool, technology, or procedure that is implemented or used in the development, maintenance, or optimization of software systems \cite{keele2007guidelines}.
In our mapping study, we did not limit our research to a single or specific intervention within the field of quantum software engineering. Instead, we sought to broadly examine the various interventions implemented or proposed within this field. This decision allowed us to gain a more comprehensive understanding of the current state of research in quantum software engineering and identify potential trends or gaps in the existing literature.
By aggregating the keywords from the P and I and the research questions, we formulated the following research string:

\vspace{5pt} %
    \noindent\fcolorbox{black}{blue!03}{%
        \parbox{0.97\linewidth}{% 
            ("quantum" AND "software" AND ("engineering" OR "development" OR "requirements" OR "quality" OR "design*" OR "test*" OR "maintenance" OR "management"))
        }%
    }%
    \vspace{5pt} %

The guidelines \slrguidelines suggest that employing IEEE and ACM, in addition to two indexing databases, would be sufficient for our mapping study. 
Therefore, we selected ACM Digital Library, IEEE Xplore, SCOPUS, and Web of Science.
We applied our search query to each database, examining all fields.
\Cref{tab:studies} depicts the number of studies obtained by applying the query to each database.

\subsection{Study Selection and Quality Assessment}

\begin{figure*}

    \includegraphics[width=\columnwidth]{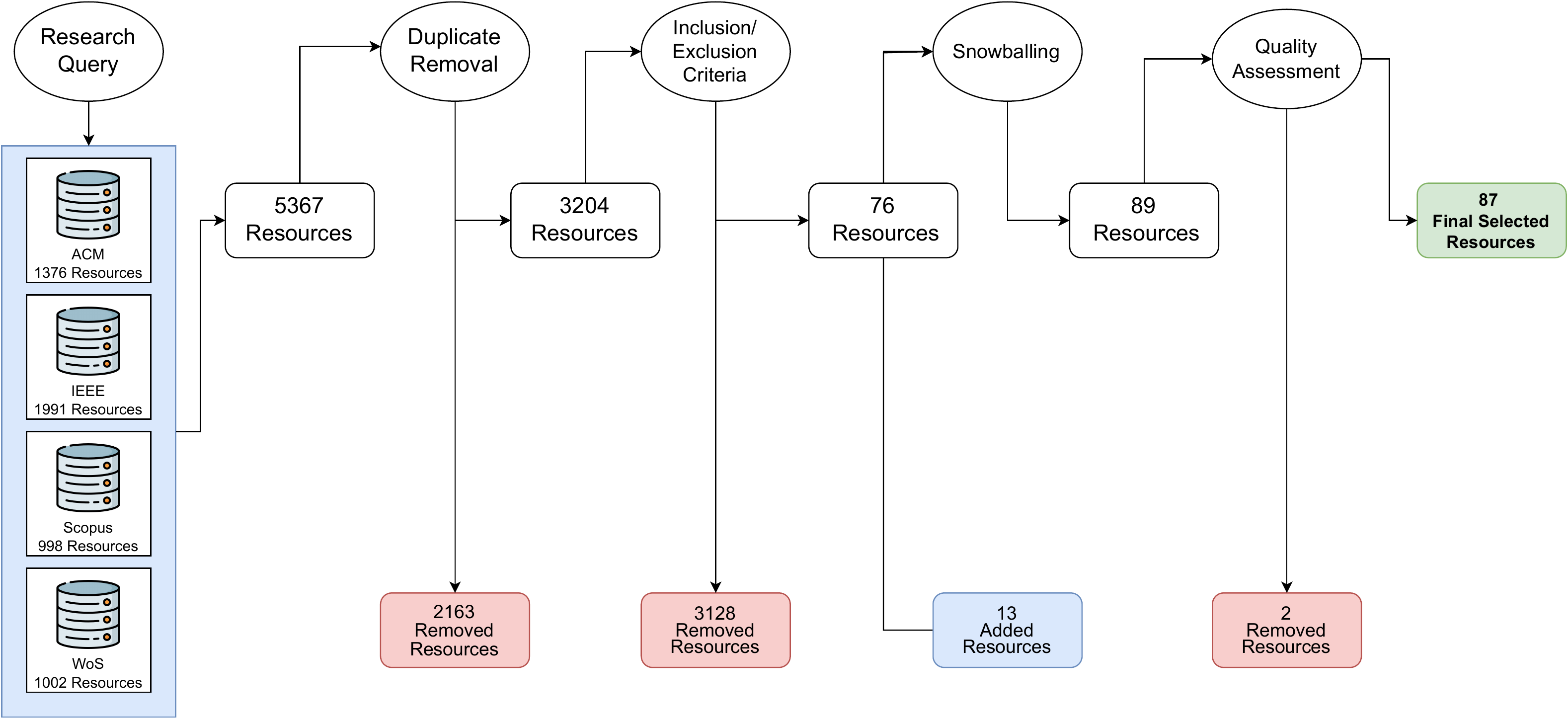}
    
    \caption{Representation of the study selection process.
    From the initial set of 3,204 non-duplicated primary studies, 3,128 articles were removed by applying the criteria on title and abstract initially, and on the full text afterward.
    By applying the snowballing, 13 articles were added (already filtered with inclusion and exclusion criteria).
    Finally, two articles were removed by applying the quality assessment criteria.
    Hence, the final set of included primary studies counts 87 articles.
    }
\end{figure*}

% Research Method   & Research method as reported in the guidelines \cite{petersen2015guidelines}.                                                            & RQ$_{1.2}$ \\

The relevant papers were selected considering the title, abstract, and full-text reading.
The inclusion and exclusion criteria were specified and discussed among all the authors \cite{petersen2015guidelines}.
We applied a think-aloud practice describing the process of inclusion/exclusion to one study to align understanding.
Based on this discussion, the level of agreement was determined, and the criteria were updated accordingly.
After this, the final criteria were applied to the studies by the first author.
The third author then reviewed the whole selection process.
Articles satisfying at least one of the exclusion criteria listed in \Cref{tab:study_selection_criteria} were excluded.
Consequently, articles that were not already excluded by meeting an exclusion criterion and satisfied at least one of the inclusion criteria listed in \Cref{tab:study_selection_criteria} were included.

\begin{table}[!t]
    \centering
    \caption{Study selection criteria applied in the study selection phase.}
    \label{tab:study_selection_criteria}
    \begin{tabular}{@{}p{0.4\columnwidth}@{\hskip 0.5in}p{0.4\columnwidth}@{}}
    \toprule
    \multicolumn{2}{c}{\textbf{Study Selection Criteria}} \\ \midrule
    \multicolumn{1}{c}{\textbf{Exclusion Criteria}} & \multicolumn{1}{c}{\textbf{Inclusion Criteria}} \\\midrule
    Article not peer-reviewed. & Peer-reviewed article in English published not earlier than 2018. \\
    Article not in English. & Article discussing or proposing SE techniques or practices applied to quantum software development or the quantum software lifecycle. \\
    Article published before 2018. & Article discussing an empirical investigation on some socio-technical aspects of software engineering applied to quantum software development. \\
    Conference articles extended in journals. & \\
    Conference summary. & \\
    Secondary or tertiary study on the matter. & \\
    Article not already selected. & \\ \bottomrule
    \end{tabular}
\end{table}

Applying the whole approach resulted in a total of 76 selected resources.
As complementary research approach \cite{petersen2015guidelines, wohlin2013reliability}, we performed both forward and backward snowballing ---\ie the selection of publications that cite (\textit{forward snowballing}), or are cited (\textit{backward snowballing}) by each selected study--- until no other unseen publication could be added, to guarantee the maximum coverage possible.
After filtering them with the same selection criteria (\Cref{tab:study_selection_criteria}), a total of 13 resources were added by the end of this procedure.

\begin{table}[!t]
\renewcommand{\arraystretch}{1.7}
\centering
\caption{Essential data elements collected from the resources, including study ID, article title, research type, research method, SWEBOK focus area, result type, main tool/framework, publication year, authors, and venue/venue type.}\label{tab:data-form}
\begin{tabular}{@{}lp{0.6\columnwidth}l@{}}
\toprule
Data Item         & Description                                                                                                     & RQ         \\ \midrule
Study ID          & Integer                                                                                                   &            \\
Article Title     & Name of the article.                                                                                      &            \\
Research Type     & Type of research as reported in the guidelines \cite{petersen2008systematic,petersen2015guidelines}. A detailed description is given in \Cref{tab:research_type}                                                          & RQ$_{1.2}$, RQ$_{1.3}$ \\
SWEBOK KA         & SWEBOK Knowledge Area \cite{bourque2004swebok} as representative of a SE topic. A detailed description is given in \Cref{tab:knowledge_areas}                                         & RQ$_{1.1}$, RQ$_{1.3}$ \\
Result Type       & Type of result achieved by the article. Classification depicted in \Cref{tab:res-types}. & RQ$_{2.1}$,  RQ$_{2.2}$ \\
Tool/Framework    & Main quantum tool or framework that was the object of study in the article.                                   & RQ$_{2.2}$ \\
Year              & Publication year of the article.                                                                          & RQ$_{3.1}$, RQ$_{3.2}$ \\
Authors           & List of authors of the article.                                                                           & RQ$_{4.1}$, RQ$_{4.2}$\\
Venue             & Venue of publication of the article.                                                                      & RQ$_{5.1}$ \\
Venue Type        & Type of the publication venue, as depicted in the guidelines (Journal, Conference, Book, Magazine) \cite{petersen2015guidelines}.                                             & RQ$_{5.1}$ \\ \bottomrule
\end{tabular}
\end{table}

\begin{table}
	\centering
	\caption{SWEBOK Knowledge Areas as described in literature \cite{bourque2004swebok}}
	\label{tab:knowledge_areas}
        \begin{adjustbox}{max width=.98\columnwidth}
	\begin{tabular}{lp{0.8\textwidth}}
        \toprule
        Knowledge Area                             & Description\\
        \midrule
		Software Requirements                      &  The Software Requirements Knowledge Area deals with gathering, analyzing, specifying, and validating the requirements of a software product. These activities are crucial for the success of software engineering projects and ensure that the product meets the needs and constraints of the real-world problems it is intended to solve.\\
  
		Software Design                            & The Software Design KA involves defining a software system's architecture, components, and behavior and creating a detailed plan for its construction based on software requirements analysis. The resulting software design provides a comprehensive blueprint for the software's construction.\\
  
		Software Construction                      & The Software Construction Knowledge Area involves detailed software creation through design, coding, testing, debugging, and verification to satisfy requirements and design constraints. It covers software construction fundamentals, management, technologies, practical considerations, and tools.\\
  
		Software Testing                           & The Software Testing Knowledge Area involves evaluating and improving software quality by identifying defects through dynamic verification against expected behavior on a finite set of test cases. It covers software testing fundamentals, techniques, user interface testing and evaluation, test-related measures, and practical considerations. \\
  
		Software Maintenance                       & The Software Maintenance Knowledge Area covers enhancing, adapting, and correcting existing software through perfective, adaptive, and corrective maintenance. It includes maintenance fundamentals such as categories and costs, key technical and management issues, cost estimation, and measurement. It also covers the maintenance process of software maintenance techniques such as program comprehension, re-engineering, reverse engineering, refactoring, software retirement, disaster recovery techniques, and maintenance tools.\\
  
        Software Configuration Management          &  The Software Configuration Management Knowledge Area deals with identifying and controlling the configuration of software systems at different points in time and maintaining the integrity and traceability of the configuration throughout the software life cycle. It includes managing the SCM process, software configuration identification, control, status accounting and auditing, software release management and delivery, and software configuration management tools.\\
        
        Software Engineering Management            & The Software Engineering Management Knowledge Area involves planning, coordinating, measuring, reporting, and controlling software projects and programs to ensure systematic, disciplined, and quantified development and maintenance. It covers initiation and scope definition, software project planning including process planning, effort estimation, risk analysis, and quality planning, software project enactments such as measuring, reporting, controlling, acquisition, and supplier contract management, product acceptance, review and analysis of project performance, project closure, and software management tools.\\
        
		Software Engineering Process               & The Software Engineering Process Knowledge Area deals with defining, implementing, assessing, measuring, managing, and improving software life cycle processes. It covers process implementation and change, process definition, software life cycle models and processes, process adaptation, and process automation. It also includes process assessment models and methods, measurement of software processes and products, measurement techniques, quality of measurement results, and software process tools.\\

        Software Engineering Models and Methods    & This KA covers modeling principles, types, and analysis for correctness, completeness, consistency, quality, and tradeoffs. It also covers software development methods like heuristic, formal, prototyping, and agile, but specific methods for particular life cycle stages are covered elsewhere.\\
        
        Software Quality                           & Software quality is an important aspect of software development and is covered in various SWEBOK V3 KAs. The Software Quality KA covers software quality fundamentals, quality management processes, and practical considerations such as defect characterization and measurement, as well as software quality tools.\\
        
        Software Engineering Professional Practice & Software engineering professional practice involves the professional and ethical practice of software engineering, including professionalism, codes of ethics, group dynamics, and communication skills. The Software Engineering Professional Practice KA covers topics such as professional conduct, software engineering standards, legal issues, working in teams, interacting with stakeholders, and dealing with multicultural environments. \\
		\bottomrule
	\end{tabular}
        \end{adjustbox}
\end{table}

\begin{table}
	\centering
	\caption{Research type facet as depicted in literature \cite{petersen2008systematic,petersen2015guidelines, wieringa2006requirements}}
	\label{tab:research_type}
        \begin{adjustbox}{max width=.98\columnwidth}
	\begin{tabular}{lp{0.8\textwidth}}
        \toprule
        Category                             & Description\\
        \midrule
		 Validation Research & Methods examined are unique and have not yet been used in practice. Experiments, or laboratory work, are an example.\\
        Evaluation Research & Techniques are put into practice, and their effectiveness is assessed. That is, it is demonstrated how the technique is applied in practice (solution implementation) and what the consequences of the application are in terms of benefits and drawbacks (implementation evaluation). This also includes identifying industry problems.\\
        Solution Proposal & A problem solution is proposed; the solution can be either novel or a significant extension of an existing technique. A small example or a good line of argumentation demonstrates the potential benefits and applicability of the solution.\\
        Philosophical Paper & This kind of paper proposes a new way of viewing existing things by organizing the field into a taxonomy or conceptual framework.\\
        Opinion Paper & This kind of paper expresses someone's personal opinion on whether a particular technique is good or bad, or how things should be done. They make no use of related work or research methodologies.\\
        Experience Paper & This kind of paper is written to describe the practical details of what was done and how it was accomplished, based on the personal experience of the author.\\
     \bottomrule
	\end{tabular}
        \end{adjustbox}
\end{table}

The final step of the selection phase consists of assessing the quality of the extracted resources to limit the risk of bias and incorrect results \slrguidelines.
Despite Kitchenam \etal \cite{keele2007guidelines} do not require systematic mapping studies to conduct a critical appraisal procedure, Petersen \etal \cite{petersen2015guidelines} recommend conducting a quality assessment but not pose too high requirements, to not exclude resources that might be relevant in the scope of a systematic mapping study.
Hence, we developed the following quality assessment strategy.
We defined a checklist that comprises the following questions:
\begin{enumerate}
    \item Is the motivation of the paper clearly and explicitly reported?
    \item Is the main outcome of the paper clearly and explicitly reported?
\end{enumerate}

The following scores are assigned to each of the above questions.
1 if the answer to the question is \textit{Yes, explicitly} , 0.5 if the answer is \textit{Yes, but not explicitly reported}, 0 if the answer is \textit{Not Reported}.
The score of both questions is then summed up.
A primary study that achieves a minimum score of 1 is considered acceptable.
Applying this quality assessment procedure led to the exclusion of two articles.
To conclude, applying the whole procedure resulted in the acceptance of 87 primary studies.

\subsection{Data Collection and Analysis}

\begin{table}[!t]
\caption{Classification of the result types.}\label{tab:res-types}
\begin{tabular}{@{}lp{0.80\textwidth}@{}}
\toprule
Result Type & Description                                                                                                             \\ \midrule
Empirical   & These results generally, but not exclusively, come out from an empirical study. The result is not directly exploitable but sheds light on a phenomenon. This kind of result might come from both Validation and Evaluation research and from Experience, Opinion, and Philosophical papers. \\
Technique   & This result comes from a technique to solve a particular issue. The technique is validated through an empirical study.           \\
Tool        & This result has the same characteristics of a technique but is released as an available tool.                   \\
Dataset     & This result is a collection of useful data or similar. It can be exploited as a reference or for empirical studies. \\ 
Guidelines & This type of result is a collection of suggestions for a particular scenario, \eg for developing a cloud-based quantum hybrid application.\\
Catalog & Similar to a dataset, but more abstract, it is a collection of useful references (\eg data encoding patterns).\\
Position & This result concerns publications that provide no explicit result but formulate a position or opinion, such as philosophical papers.\\\bottomrule
\end{tabular}
\end{table}

We created an extraction form shown in \Cref{tab:data-form} to gather information from the primary studies. 
This template includes various data items and their corresponding values.
It is possible to note that the rationale behind the choice of this gathered information is pretty straightforward: each of the extracted data items corresponds to the phenomenon each research question aims to solve.
\Cref{tab:data-form} clearly explains this in the description column.
The process of extracting this data involved the first author collecting the information and the third author reviewing it for accuracy by comparing it to the original statements in each paper; this was done to mitigate the risk of subjectivity, as depicted in the guidelines \slrguidelines.

Petersen \etal \cite{petersen2015guidelines} recommend using topic-independent classifications as much as possible.
This classification was conducted for the SWEBOK Knowledge Areas \cite{bourque2004swebok} (depicted in \Cref{tab:knowledge_areas}), the research type (depicted in \Cref{tab:research_type}) \cite{wieringa2006requirements, petersen2008systematic, petersen2015guidelines}, and the venue \cite{petersen2015guidelines}.
Nonetheless, for the result type, a topic-specific classification was needed.
Hence, as suggested by Petersen \etal \cite{petersen2015guidelines}, we used open coding \cite{strauss1994grounded} to organize the articles into categories and then counted the number of articles in each category.
We labeled or assigned keywords to concepts found in the text during the open coding process, resulting in several open codes, which were then organized into a larger structure. 
In this process, we merged or renamed some of the codes representing categories \cite{strauss1994grounded}. 
Then, we used these categories to classify the papers.
We applied this process to the abstracts of the articles.
However, if the abstracts were not clear enough, we considered other parts of the paper (\eg introduction and conclusion).
The resulting classification is depicted in \Cref{tab:res-types}.

We employed descriptive statistics tools to analyze the gathered data, as suggested by the guidelines, \slrguidelines.
In particular, we relied on tables (with bibliographical references) to report absolute frequencies of the categorical variables (\eg the SWEBOK Knowledge Area of interest of each resource) or bar charts (\eg absolute number of papers per year).
To represent co-occurrences of categorical variables, we employed heatmaps and pivot tables depicted as heatmaps (\eg Knowledge Areas by Research Types).
To show the evolution over time of the number of published resources, we employed line graphs (\eg cumulative number of resources per publication year).

\subsection{Threats to Validity}
The following discusses potential threats and limitations to the study's validity. 

\paragraph{Descriptive Validity}
The extent to which observations are accurately and objectively described is descriptive validity.
Threats to descriptive validity are prevalent in qualitative and quantitative studies.
As reported by the guidelines~\slrguidelines, a data collection form has been designed to support data recording to reduce this threat.
As with the primary studies, the form objectified the data extraction process and allowed it to be revisited regularly.
Moreover, as depicted by the guidelines \cite{petersen2015guidelines}, not only we extracted the research type and method (which were also fundamental for answering our research questions), but we also mainly applied topic-independent classifications.
For topic-specific classification, \ie for the main results of the primary studies, we applied open coding \cite{strauss1994grounded}, as the guidelines recommend \cite{petersen2015guidelines}.
As a result, this threat is considered to be under control.

\paragraph{Theoretical Validity}
The ability to capture what we intend to capture determines theoretical validity. 
Furthermore, confounding factors such as biases and subject selection play a significant role.
In our case, this refers to the study identification (or sampling) and the data extraction and classification.

As Wohlin \etal \cite{wohlin2013reliability} pointed out, the selection of the studies could have threatened the validity of the study.
To mitigate this risk, we supplemented the search with backward and forward snowball sampling \cite{jalali2012systematic} of all studies included by full-text reading.
In addition, to increase the reliability of the inclusion and exclusion criteria, we applied the think-aloud protocol to the criteria defined and discussed among all the authors.
Then the first author conducted the selection process that was reviewed by the third afterward.

Researcher bias is a threat regarding data extraction and classification. Following the guidelines \slrguidelines, the third author assessed all extractions made by the first author to mitigate this threat. 
Nonetheless, since this step involves human judgment, the threat cannot be entirely excluded.

\begin{table}
\centering
\caption{Number of resources per Swebok Knowledge Area. The references are listed in the Bibliography.}
\label{tab:rq1_1}
\begin{tabular}{lrp{0.3\textwidth}}
\toprule
                     Swebok Knowledge Area & Resources &                                                                                                                                                                                                                                                                                                                                                                  References \\
\midrule
                          Software Testing &        22 & \citeR{Fortunato2022-io, Fortunato2022-or, Pontolillo2022-ep, Abreu2022-mm, Trinca2022-ku, Wang2022-es, Burgholzer2021-kf, Mykhailova2021-ss, Li2020-qe, Zhou2019-gi, Huang2019-de, M_Paltenghi2022-rl, X_Wang2021-wy, E_Mendiluze2021-ls, S_Ali2021-gl, A_Miranskyy2019-ab, X_Wang2021-ew, J_Wang2021-jl, J_Wang2021-fg, Honarvar2020-gj, Wang2022-ua, wang2021generating} \\
                     Software Construction &        17 &                                                                                 \citeR{Hevia2022-gw, Moguel2022-sc, Da_Rosa2022-nq, Gheorghe-Pop2022-vh, Humble2021-ol, Sodhi2021-er, Vietz2021-cx, Kruger2020-rc, Hevia_Oliver2020-le, Gomes2020-mw, Dreher2019-qw, LaRose2019-ti, Fingerhuth2018-gm, Steiger2018-kg, J_Garcia-Alonso2022-nv, Cobb2022-zl, Serrano2022-de} \\
   Software Engineering Models and Methods &        15 &                                                                                          \citeR{Alonso2022-ye, Perez-Castillo2021-jz, Gemeinhardt2021-kt, Scheerer2021-rw, Piattini2021-mg, Perez-Delgado2020-ed, Jimenez-Navajas2020-ew, Moguel2020-yc, Barbosa2020-we, Ali2020-wa, Fu2021-jz, Jimenez-Navajas2021-eb, Kumara2021-ox, Perez-Castillo2022-kx, Weder2021-cf} \\
                           Software Design &        10 &                                                                                                                                                                                                 \citeR{Exman2022-gm, Schonberger2022-qr, Exman2021-yg, N_Oldfield2022-lo, Weigold2022-lw, Leymann2019-ot, Weigold2021-tb, M_Weigold2021-hq, Sanchez2021-oo, Weigold2021-nv} \\
                          Software Quality &         9 &                                                                                                                                                                                                                        \citeR{LaRose2022-wt, Campos2021-ri, Zhao2021-gx, Verduro2021-nk, Saraiva2021-cm, P_Zhao2021-ou, P_Zhao2021-iq, Paltenghi2022-ll, Cruz-Lemus2021-it} \\
                      Software Maintenance &         7 &                                                                                                                                                                                                                          \citeR{Openja2022-lt, Perez-Castillo2022-lq, Luo2022-gp, Perez-Castillo2021-zs, Miranskyy2020-mp, R_Perez-Castillo2022-ms, Jimenez-Navajas2020-jm} \\
Software Engineering Professional Practice &         4 &                                                                                                                                                                                                                                                                                                  \citeR{De_Stefano2022-jz, Hughes2022-rr, El_Aoun2021-lu, Peterssen2020-yr} \\
             Software Engineering Process  &         3 &                                                                                                                                                                                                                                                                                                                 \citeR{L_Nallamothula2020-md, B_Weder2021-nt, Weder2020-ie} \\
\bottomrule
\end{tabular}
\end{table}

\begin{table}
\centering
\caption{Resources by research type. The references are listed in the bibliography.}
\label{tab:rq1_2}
\begin{tabular}{lrp{0.5\columnwidth}}
\toprule
       Research Type & Resources &                                                                                                                                                                                                                                                                                                                                                                                                                                                                                                                                                                 References \\
\midrule
   Solution Proposal &        32 & \citeR{Alonso2022-ye, Fortunato2022-io, Perez-Castillo2022-lq, LaRose2022-wt, Fortunato2022-or, Abreu2022-mm, Wang2022-es, Da_Rosa2022-nq, Perez-Castillo2021-jz, Zhao2021-gx, Burgholzer2021-kf, Perez-Castillo2021-zs, Dreher2019-qw, Zhou2019-gi, Huang2019-de, Steiger2018-kg, X_Wang2021-wy, R_Perez-Castillo2022-ms, E_Mendiluze2021-ls, S_Ali2021-gl, X_Wang2021-ew, P_Zhao2021-ou, J_Wang2021-jl, J_Garcia-Alonso2022-nv, J_Wang2021-fg, Honarvar2020-gj, Wang2022-ua, wang2021generating, Fu2021-jz, Jimenez-Navajas2021-eb, Perez-Castillo2022-kx, Weder2021-cf} \\
Philosophical Papers &        22 &                                                                                                                                                               \citeR{Gheorghe-Pop2022-vh, Gemeinhardt2021-kt, Scheerer2021-rw, Vietz2021-cx, Piattini2021-mg, Perez-Delgado2020-ed, Hevia_Oliver2020-le, Jimenez-Navajas2020-ew, Gomes2020-mw, M_Paltenghi2022-rl, N_Oldfield2022-lo, L_Nallamothula2020-md, B_Weder2021-nt, Weder2020-ie, Ali2020-wa, Leymann2019-ot, Cruz-Lemus2021-it, Weigold2021-tb, Kumara2021-ox, M_Weigold2021-hq, Sanchez2021-oo, Weigold2021-nv} \\
 Validation Research &        15 &                                                                                                                                                                                                                                                                                                                            \citeR{Openja2022-lt, De_Stefano2022-jz, Exman2022-gm, Luo2022-gp, Pontolillo2022-ep, Trinca2022-ku, Schonberger2022-qr, Exman2021-yg, Campos2021-ri, Sodhi2021-er, El_Aoun2021-lu, Li2020-qe, Kruger2020-rc, Paltenghi2022-ll, Serrano2022-de} \\
   Experience Papers &        11 &                                                                                                                                                                                                                                                                                                                                                                                   \citeR{Hevia2022-gw, Moguel2022-sc, Mykhailova2021-ss, Saraiva2021-cm, Miranskyy2020-mp, Peterssen2020-yr, LaRose2019-ti, Fingerhuth2018-gm, P_Zhao2021-iq, Weigold2022-lw, Cobb2022-zl} \\
      Opinion Papers &         5 &                                                                                                                                                                                                                                                                                                                                                                                                                                                                                   \citeR{Humble2021-ol, Verduro2021-nk, Moguel2020-yc, A_Miranskyy2019-ab, Barbosa2020-we} \\
 Evaluation Research &         1 &                                                                                                                                                                                                                                                                                                                                                                                                                                                                                                                                                      \citeR{Hughes2022-rr} \\
\bottomrule
\end{tabular}
\end{table}

\section{Analysis of the Results} \label{sec:results}

The following section presents the results of our mapping study.
Each of the following paragraphs aims to answer the main topics posed by our research questions.

\subsection{RQ1: Main Topics and Studies in QSE}

\begin{figure*}[!ht]
    \centering
    \includegraphics[width=0.8\textwidth, height=!]{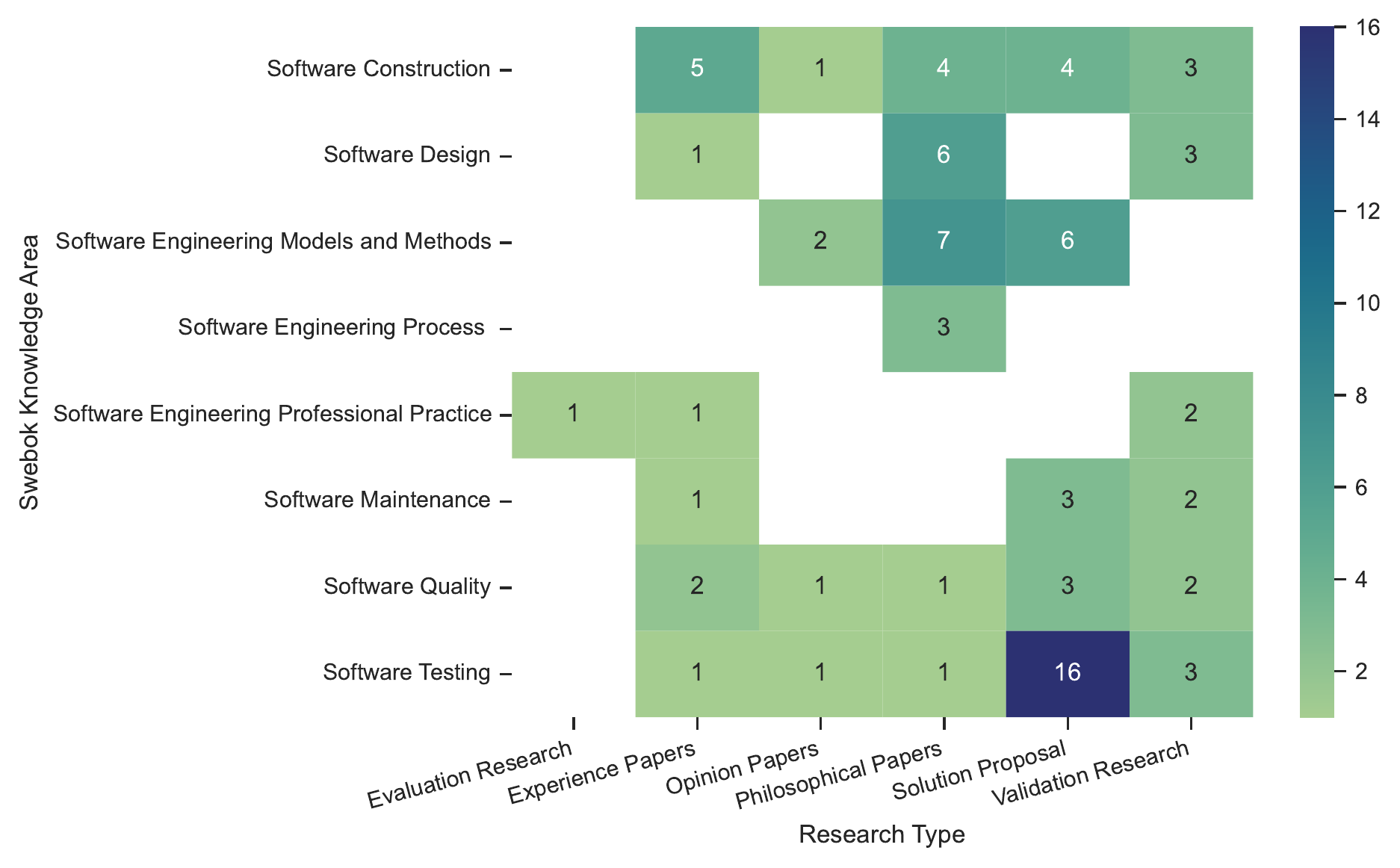}
    \caption{Distribution of published papers by knowledge area and research type. The number of papers is represented by color darkness, with darker colors indicating a higher number of papers. Most papers focus on software testing, followed by software engineering models, methods, and construction. Solution proposals are the most common type of published paper, followed by philosophical papers. Empirical studies, represented by validation and evaluation research categories, are relatively rare.}
    \label{fig:rq1_1}
\end{figure*}

\Cref{tab:rq1_2} depicts the distribution of the analyzed resources for the employed research type.
It is evident that Solution Proposal papers are the most common type of research, with a total of 32 papers across all focus areas.
Philosophical papers are the second most common type of research, with a total of 22 resources.
On the other hand, the Evaluation Research and Validation Research categories (which denote industrial and laboratory empirical studies, respectively) have fewer resources ( 1 and 15, respectively).
Experience papers are also quite numerous (15), whilst Opinion papers appear in a small number (5).

Regarding the knowledge areas, whose distribution can be seen in \Cref{tab:rq1_1}, it can be noted that some focus areas have a higher number of publications than others. 
For instance, the focus area of Software Testing has the highest number of publications, with 22 papers across all research types. 
On the other hand, the software engineering process is the least investigated knowledge area, with just three resources.
The other two most investigated knowledge areas are Software Construction and Software Engineering Models and Methods, with 17 and 15 resources each.
The other knowledge areas present 10 resources or less.
It is worth mentioning that several knowledge areas, such as software requirements, configuration management, and software engineering management, have not been the subject of any published papers.

\Cref{fig:rq1_1} gives insights into the variation in the distribution of research types across the different focus areas.
In the figure, the number of papers is represented by color darkness, with darker colors indicating a higher number.
It is possible to see that Philosophical papers focus on Software Design, Software Engineering Models and Methods, and Software Construction, with six, seven, and four resources, respectively, but they are almost nonexistent in the other knowledge areas. 
On the other hand, validation research resources are more evenly distributed over all the knowledge areas, with two to three resources each, except for Software Engineering Models and Methods and Software Engineering Process, which have none.
This phenomenon of some focus areas having a relatively even distribution of papers across the different research types while others have a higher concentration of specific papers can also be observed in other cases. 
For example, the focus area of Software Quality has a relatively even distribution of papers across the different research types, with one to three resources for each of the research types considered (except for evaluation research). 
In contrast, the focus area of Software Testing has a higher concentration of Solution Proposal papers (16) and very few resources in the other research categories. 

\begin{summarybox}[\textbf{RQ$_{1}$}]
Research in QSE is dominated by Solution Proposal and Philosophical papers, with Empirical studies receiving slightly less attention overall (\textbf{RQ$_{1.1}$}). 
While Software Testing, Construction, and Models and Methods receive significant attention, other areas, such as the Software Engineering Process and Management areas have been largely neglected (\textbf{RQ$_{1.2}$}). 
The distribution of research types across focus areas is uneven, with some areas having a higher concentration of specific papers than others. 
For example, Solution Proposal papers are abundant in Software Testing, while Validation Research resources are more evenly distributed across most areas, with some exceptions (\textbf{RQ$_{1.3}$}).
\end{summarybox}

\subsection{RQ2: Main results obtained in QSE}

\begin{table}[!ht]
    \centering
    \caption{Resources per result type.}\label{tab:rq2_1}
    \begin{tabular}{lrp{0.5\columnwidth}}
\toprule
Result Type & Resources &                                                                                                                                                                                                                                                                                                                                                                                                 References \\
\midrule
  Technique &        22 & \citeR{Abreu2022-mm, Perez-Castillo2021-jz, Scheerer2021-rw, Li2020-qe, Huang2019-de, Jimenez-Navajas2021-eb, Kumara2021-ox, Perez-Castillo2022-kx, Alonso2022-ye, Zhou2019-gi, X_Wang2021-wy, Wang2022-ua, wang2021generating, Perez-Castillo2021-zs, S_Ali2021-gl, J_Wang2021-jl, J_Wang2021-fg, Honarvar2020-gj, Jimenez-Navajas2020-jm, Weder2021-cf, R_Perez-Castillo2022-ms, J_Garcia-Alonso2022-nv} \\
  Empirical &        21 &                                             \citeR{Openja2022-lt, De_Stefano2022-jz, Luo2022-gp, Pontolillo2022-ep, Moguel2022-sc, Schonberger2022-qr, Campos2021-ri, Sodhi2021-er, El_Aoun2021-lu, Paltenghi2022-ll, Exman2022-gm, Hughes2022-rr, Exman2021-yg, Gemeinhardt2021-kt, Vietz2021-cx, Saraiva2021-cm, Perez-Delgado2020-ed, N_Oldfield2022-lo, B_Weder2021-nt, Serrano2022-de, Trinca2022-ku} \\
   Position &        21 &                     \citeR{Humble2021-ol, Verduro2021-nk, Piattini2021-mg, Hevia_Oliver2020-le, Jimenez-Navajas2020-ew, Gomes2020-mw, Moguel2020-yc, Peterssen2020-yr, M_Paltenghi2022-rl, L_Nallamothula2020-md, A_Miranskyy2019-ab, Barbosa2020-we, Weder2020-ie, Sanchez2021-oo, LaRose2019-ti, Fingerhuth2018-gm, Ali2020-wa, Kruger2020-rc, Miranskyy2020-mp, Mykhailova2021-ss, Gheorghe-Pop2022-vh} \\
       Tool &        13 &                                                                                                                                                                                \citeR{Fortunato2022-io, Fortunato2022-or, Wang2022-es, Dreher2019-qw, E_Mendiluze2021-ls, X_Wang2021-ew, Hevia2022-gw, LaRose2022-wt, Steiger2018-kg, Da_Rosa2022-nq, Burgholzer2021-kf, Perez-Castillo2022-lq, Fu2021-jz} \\
    Catalog &         7 &                                                                                                                                                                                                                                                                                   \citeR{Zhao2021-gx, Weigold2022-lw, Cruz-Lemus2021-it, Weigold2021-tb, M_Weigold2021-hq, Weigold2021-nv, Leymann2019-ot} \\
    Dataset &         2 &                                                                                                                                                                                                                                                                                                                                                                       \citeR{P_Zhao2021-ou, P_Zhao2021-iq} \\
 Guidelines &         1 &                                                                                                                                                                                                                                                                                                                                                                                        \citeR{Cobb2022-zl} \\
\bottomrule
\end{tabular}

\end{table}

\begin{table}[!ht]
    \centering
    \caption{Resources per quantum technology.}\label{tab:rq2_2}
    \begin{tabular}{lrp{0.5\columnwidth}}
\toprule
Quantum Technology & Resources &                                                                                                                                                                                                                                                                                                                                                                                                                                                                                                                                                                                                                                                                                                              References \\
\midrule
              None &        40 & \citeR{Humble2021-ol, Verduro2021-nk, Piattini2021-mg, Hevia_Oliver2020-le, Jimenez-Navajas2020-ew, Gomes2020-mw, Moguel2020-yc, Peterssen2020-yr, M_Paltenghi2022-rl, L_Nallamothula2020-md, A_Miranskyy2019-ab, Barbosa2020-we, Weder2020-ie, Sanchez2021-oo, Exman2022-gm, Hughes2022-rr, Exman2021-yg, Gemeinhardt2021-kt, Vietz2021-cx, Saraiva2021-cm, Perez-Delgado2020-ed, N_Oldfield2022-lo, B_Weder2021-nt, Serrano2022-de, Abreu2022-mm, Perez-Castillo2021-jz, Scheerer2021-rw, Li2020-qe, Huang2019-de, Jimenez-Navajas2021-eb, Kumara2021-ox, Perez-Castillo2022-kx, Zhao2021-gx, Weigold2022-lw, Cruz-Lemus2021-it, Weigold2021-tb, M_Weigold2021-hq, Weigold2021-nv, Da_Rosa2022-nq, Burgholzer2021-kf} \\
          Multiple &        20 &                                                                                                                                                                                                                                                                                                                                                                                 \citeR{Openja2022-lt, De_Stefano2022-jz, Luo2022-gp, Pontolillo2022-ep, Moguel2022-sc, Schonberger2022-qr, Campos2021-ri, Sodhi2021-er, El_Aoun2021-lu, Paltenghi2022-ll, Perez-Castillo2021-zs, S_Ali2021-gl, J_Wang2021-jl, J_Wang2021-fg, LaRose2019-ti, Fingerhuth2018-gm, Ali2020-wa, Hevia2022-gw, LaRose2022-wt, Steiger2018-kg} \\
            Qiskit &        14 &                                                                                                                                                                                                                                                                                                                                                                                                                                                                                \citeR{Fortunato2022-io, Fortunato2022-or, Wang2022-es, Dreher2019-qw, E_Mendiluze2021-ls, X_Wang2021-ew, Alonso2022-ye, Zhou2019-gi, X_Wang2021-wy, Wang2022-ua, wang2021generating, P_Zhao2021-ou, P_Zhao2021-iq, Gheorghe-Pop2022-vh} \\
                Q\# &         5 &                                                                                                                                                                                                                                                                                                                                                                                                                                                                                                                                                                                                                \citeR{Honarvar2020-gj, Jimenez-Navajas2020-jm, Perez-Castillo2022-lq, Mykhailova2021-ss, Trinca2022-ku} \\
              QASM &         3 &                                                                                                                                                                                                                                                                                                                                                                                                                                                                                                                                                                                                                                                                   \citeR{Cobb2022-zl, Leymann2019-ot, Miranskyy2020-mp} \\
             Dwave &         2 &                                                                                                                                                                                                                                                                                                                                                                                                                                                                                                                                                                                                                                                                          \citeR{Kruger2020-rc, R_Perez-Castillo2022-ms} \\
     Amazon Braket &         1 &                                                                                                                                                                                                                                                                                                                                                                                                                                                                                                                                                                                                                                                                                          \citeR{J_Garcia-Alonso2022-nv} \\
            Custom &         1 &                                                                                                                                                                                                                                                                                                                                                                                                                                                                                                                                                                                                                                                                                                       \citeR{Fu2021-jz} \\
           QuantME &         1 &                                                                                                                                                                                                                                                                                                                                                                                                                                                                                                                                                                                                                                                                                                    \citeR{Weder2021-cf} \\
\bottomrule
\end{tabular}

\end{table}

\Cref{tab:rq2_1} depicts the result types achieved by the analyzed resources.
The most frequent result is \textit{Technique} with 22 resources reporting it.
\textit{Empirical} results rapidly follow, with 21 resources reporting it as the main achievement.
This result comes not only from purely empirical studies (\ie Validation or Evaluation Research) but also from Experience papers, Opinion Papers, and Philosophical Papers.
The same number of resources reports a \textit{Position} as the main result achieved. 
Mature tools are not frequent, with only 13 resources reporting it as the main result achieved.
\textit{Catalog}, \textit{Dataset}, and \textit{Guidelines} are the least commonly achieved results, with seven, two, and one resource each.

Concerning the quantum technology which the analyzed resources focus on, details can be observed in \Cref{tab:rq2_2}.
The results show that 40 resources did not focus on any specific technology, indicating that many were not interested in exploring a particular quantum technology in depth. On the other hand, 20 resources focused on multiple quantum technologies, indicating that were interested in exploring multiple quantum technologies.
Of the specific technologies focused on by the resources, Qiskit was the most popular, with 14 resources focusing on it. 
\qsharp was the second most popular, with five resources focusing on it. 
Among the remaining technologies, QASM and Dwave were the only objects of study by more than one resource (three and two, respectively).
Amazon Braket and QuantME, as well as a custom technology, were the least popular technologies, with just a single resource focusing on them.

\begin{figure}[!ht]
    \centering
    \includegraphics[width=0.8\columnwidth, height=!]{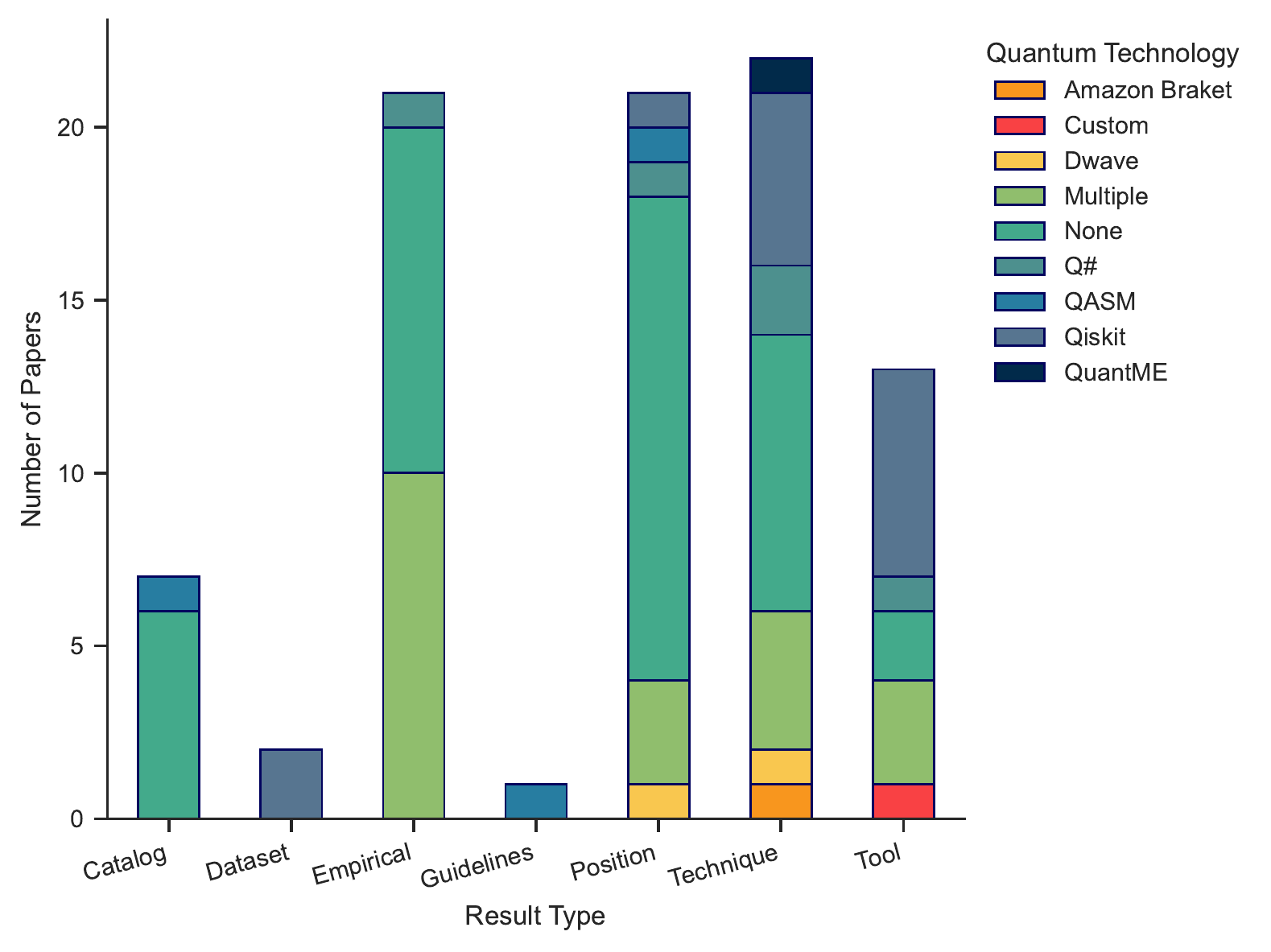}
    \caption{Distribution of the main reported result types alongside the quantum technology interested. The most frequently reported Result Type is "Technique," followed by "Multiple." Qiskit is the most commonly mentioned quantum technology, followed by "Multiple."}
    \label{fig:rq2_1}
\end{figure}

\Cref{fig:rq2_1} correlates the information carried out by \Cref{tab:rq2_1} and \Cref{tab:rq2_2} to gather further insights.
It is also possible to observe that the Result Types Empirical, Technique, Tools, and Position all showed different quantum technologies.
Specifically, empirical results mainly focused either on \textit{Multiple} or \textit{None} specific technologies, with the sole exception of a single resource focusing on \qsharp.
\textit{Position} results, on the other hand, are mainly technology independent, despite focusing on \textit{Multiple}, \qsharp, QASM, and Dwawe technologies as well.
\textit{Tool} and \textit{Technique} result types show the greatest variety of quantum technologies interested: most of the resources focusing on specific technologies fall into these categories.
\textit{Guidelines} results were only reported for QASM, while \textit{Dataset} only for Qiskit.
Finally, \textit{Catalogs} results were reported mainly in a technology-agnostic manner, except for a single resource, which focused on QASM.

\begin{summarybox}[RQ$_2$]
    Most reported results types are techniques, empirical, and positions, with the other types of results reported in fairly fewer occurrences (\textbf{RQ$_{2.1}$}).
    Most resources focus either on multiple technologies or none specific. 
    Qiskit, however, is the most commonly studied technology for techniques and tools. 
    The other technologies were rarely studied on their own (\textbf{RQ$_{2.2}$}).
\end{summarybox}

\subsection{RQ$_3$: Evolution of the discipline over time}

\begin{figure*}[!t]
    \centering
    \begin{subfigure}[t]{0.4\columnwidth}
        \centering
        \includegraphics[width=\columnwidth, height=!]{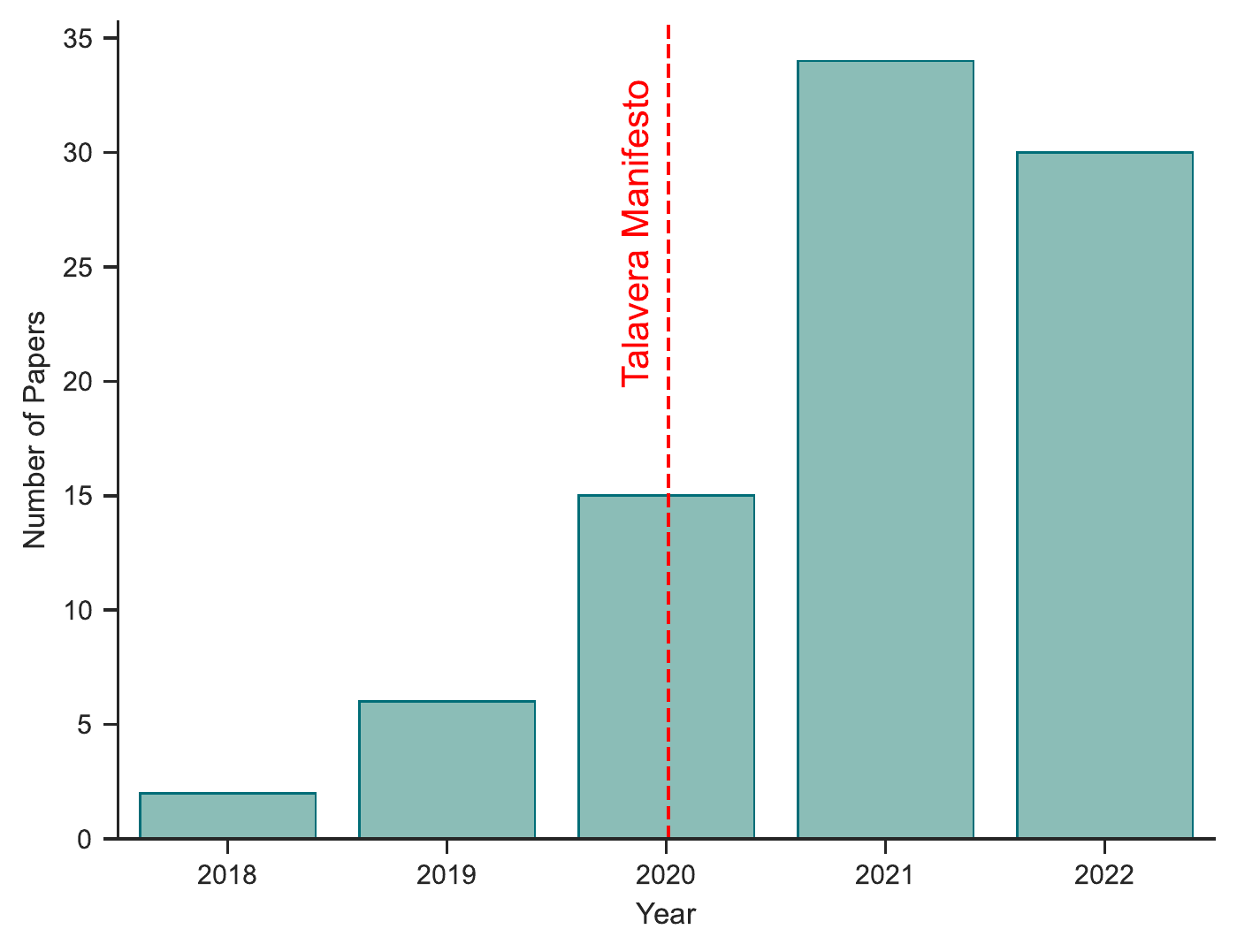}
        \caption{Publication over years}
        \label{fig:rq3_1}
    \end{subfigure}
    \begin{subfigure}[t]{0.4\columnwidth}
        \centering
        \includegraphics[width=\columnwidth, height=!]{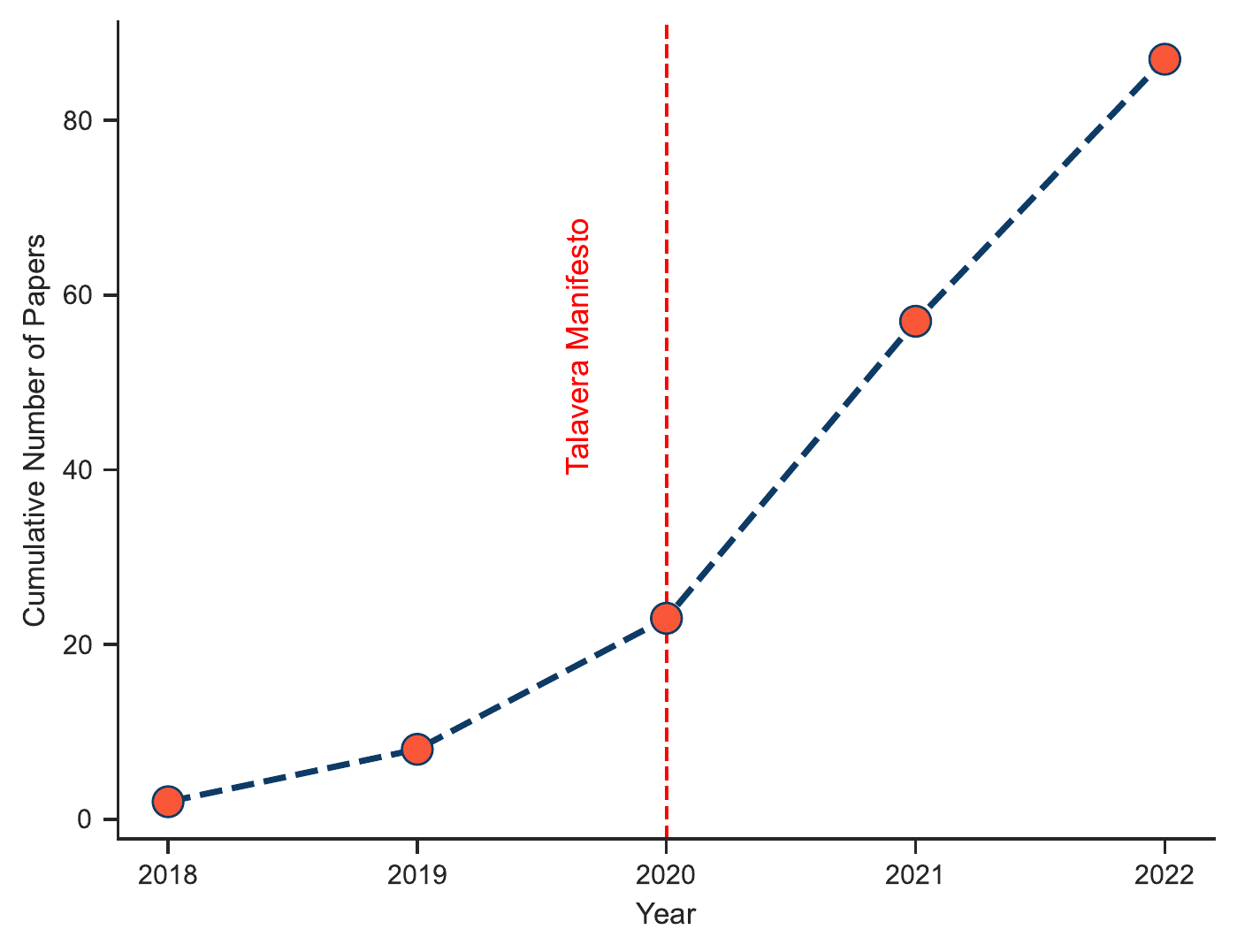}
        \caption{Cumulative number of publications over the years}
        \label{fig:rq3_2}
    \end{subfigure}
    \caption{In both graphs, the dashed line indicates the year of publication of the Talavera Manifesto, which is considered to give birth to QSE. Both graphs show a history of contributions to the quantum software engineering field before the discipline's formal establishment in 2020 and that the interest in this area among researchers has grown significantly, particularly between 2020 and 2021.}
\end{figure*}

\Cref{fig:rq3_1} depicts the number of publications over the years, while \Cref{fig:rq3_2} the cumulative number over the years.
These figures tell us that the number of papers published in this discipline has increased, with the first publications (2) dating back to 2018.
Then we have six papers in 2019, 15 in 2020, and 34 in 2021. 
The cumulative number of papers published in this field has also been increasing, with two papers published by the end of 2018, eight papers by the end of 2019, 23 papers by the end of 2020, and 57 papers by the end of 2021, reaching a total of 87 by November 2022. 

It is worth noting that the Talavera Manifesto, which established the foundations of this discipline, was published in 2020, which can roughly indicate that the field of quantum software engineering was born that year. 
However, it is interesting to observe that some papers were published in this field before the official establishment of the discipline. 
In the years following 2020, there has been a speedy increase in published papers, doubling from 14 in 2020 to 37 in 2021. 
It is also important to note that we considered the paper published up to November 2022 (the time when the research started).
Hence, data about the year 2022 might not include all the paper published in that year.

\begin{summarybox}[RQ$_3$]
    Some papers (8) were published in the field of quantum software engineering before the official establishment of the discipline in 2020 (\textbf{RQ$_{3.1}$}). The research community's interest in this discipline has been growing, with a speedy increase in published papers between 2020 and 2021 (\textbf{RQ$_{3.2}$}).
\end{summarybox}

\subsection{RQ$_4$: Authors}

\begin{figure*}[!ht]
    \centering
    \includegraphics[width=0.9\columnwidth, height=!]{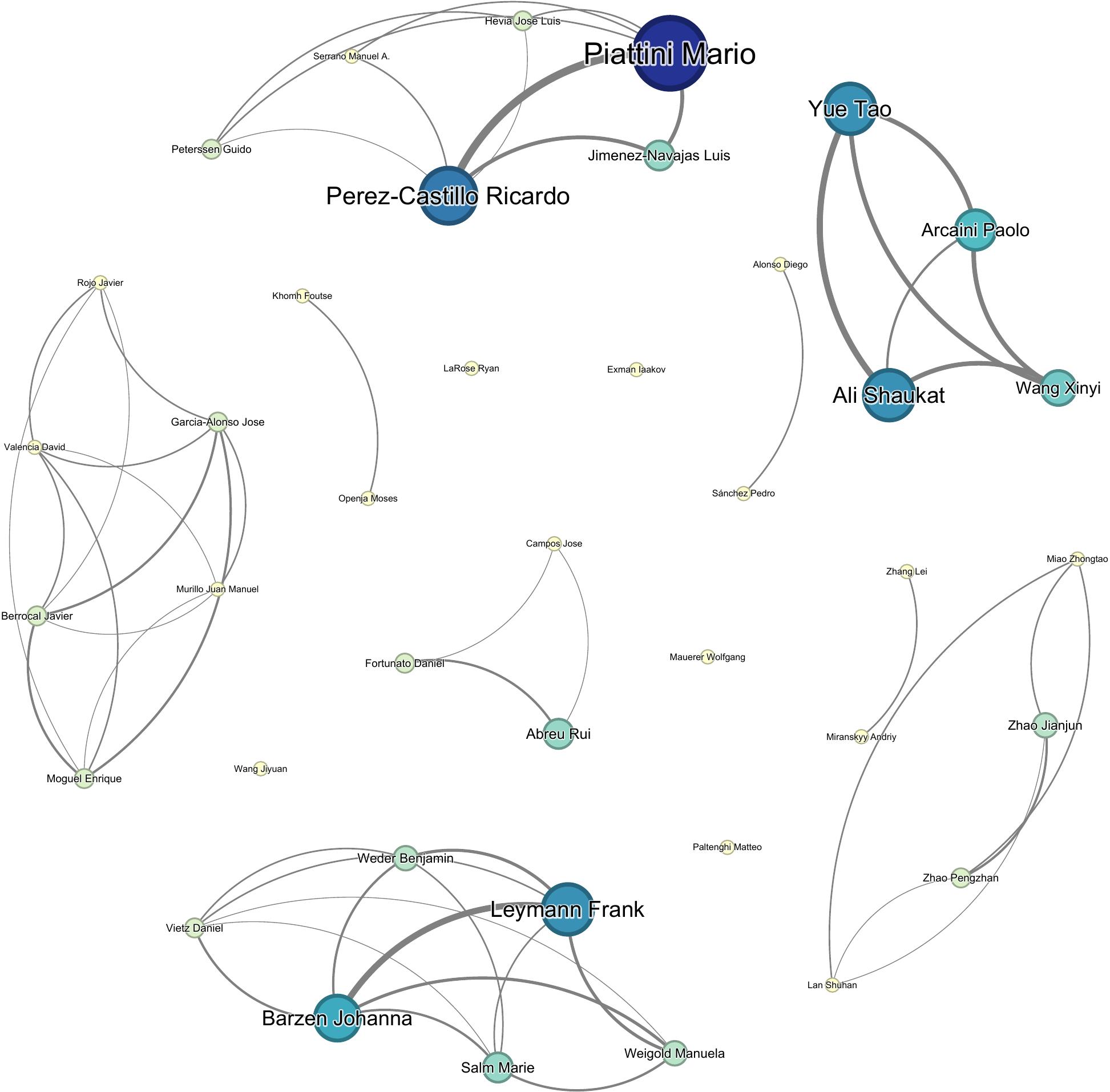}
    %\missingfigure{}
    \caption{Social network among the most productive authors in the field of Quantum Software Engineering, showing the collaborations among the authors and the number of papers they have published. The nodes' size and the edges' thickness represent the number of papers published by each author and the strength of the collaborations, respectively.}
    \label{fig:rq4_1}
\end{figure*}

\begin{table}[!ht]
      \caption{Authors by the number of published papers. Mario Piattini stands out as the most productive author, followed by Ricardo Perez-Castillo, Frank Leymann, Shaukat Ali, and Tao Yue.}
      \label{tab:rq4_1}
      \centering
      \begin{tabular}{lrp{0.5\textwidth}}
\toprule
                Author & Resources &                                                                                                                                                                                                                                                                            References \\
\midrule
        Piattini Mario &        13 & \citeR{Hevia2022-gw, Perez-Castillo2022-lq, Perez-Castillo2021-jz, Verduro2021-nk, Piattini2021-mg, Perez-Castillo2021-zs, Jimenez-Navajas2020-ew, R_Perez-Castillo2022-ms, Serrano2022-de, Jimenez-Navajas2020-jm, Jimenez-Navajas2021-eb, Cruz-Lemus2021-it, Perez-Castillo2022-kx} \\
Perez-Castillo Ricardo &        10 &                                                  \citeR{Perez-Castillo2022-lq, Perez-Castillo2021-jz, Piattini2021-mg, Perez-Castillo2021-zs, Jimenez-Navajas2020-ew, R_Perez-Castillo2022-ms, Serrano2022-de, Jimenez-Navajas2020-jm, Jimenez-Navajas2021-eb, Perez-Castillo2022-kx} \\
         Leymann Frank &         9 &                                                                                                                                    \citeR{Vietz2021-cx, B_Weder2021-nt, Weigold2022-lw, Weder2020-ie, Leymann2019-ot, Weigold2021-tb, M_Weigold2021-hq, Weder2021-cf, Weigold2021-nv} \\
           Ali Shaukat &         9 &                                                                                                                                   \citeR{Wang2022-es, X_Wang2021-wy, N_Oldfield2022-lo, E_Mendiluze2021-ls, S_Ali2021-gl, X_Wang2021-ew, Wang2022-ua, wang2021generating, Ali2020-wa} \\
               Yue Tao &         9 &                                                                                                                                   \citeR{Wang2022-es, X_Wang2021-wy, N_Oldfield2022-lo, E_Mendiluze2021-ls, S_Ali2021-gl, X_Wang2021-ew, Wang2022-ua, wang2021generating, Ali2020-wa} \\
        Barzen Johanna &         8 &                                                                                                                                                    \citeR{Vietz2021-cx, B_Weder2021-nt, Weigold2022-lw, Weder2020-ie, Weigold2021-tb, M_Weigold2021-hq, Weder2021-cf, Weigold2021-nv} \\
         Arcaini Paolo &         7 &                                                                                                                                                                  \citeR{Wang2022-es, X_Wang2021-wy, E_Mendiluze2021-ls, S_Ali2021-gl, X_Wang2021-ew, Wang2022-ua, wang2021generating} \\
            Wang Xinyi &         6 &                                                                                                                                                                                      \citeR{Wang2022-es, X_Wang2021-wy, S_Ali2021-gl, X_Wang2021-ew, Wang2022-ua, wang2021generating} \\
  Jimenez-Navajas Luis &         5 &                                                                                                                                                        \citeR{Perez-Castillo2021-jz, Jimenez-Navajas2020-ew, R_Perez-Castillo2022-ms, Jimenez-Navajas2020-jm, Jimenez-Navajas2021-eb} \\
            Salm Marie &         5 &                                                                                                                                                                                                  \citeR{Weigold2022-lw, Weder2020-ie, M_Weigold2021-hq, Weder2021-cf, Weigold2021-nv} \\
             Abreu Rui &         5 &                                                                                                                                                                                                 \citeR{Fortunato2022-io, Fortunato2022-or, Abreu2022-mm, Trinca2022-ku, Gomes2020-mw} \\
          Zhao Jianjun &         4 &                                                                                                                                                                                                                         \citeR{Luo2022-gp, Zhao2021-gx, P_Zhao2021-ou, P_Zhao2021-iq} \\
        Weder Benjamin &         4 &                                                                                                                                                                                                                      \citeR{Vietz2021-cx, B_Weder2021-nt, Weder2020-ie, Weder2021-cf} \\
       Weigold Manuela &         4 &                                                                                                                                                                                                              \citeR{Weigold2022-lw, Weigold2021-tb, M_Weigold2021-hq, Weigold2021-nv} \\
\bottomrule
\end{tabular}

\end{table}

\Cref{tab:rq4_1} depicts the authors that published more than three papers in QSE.
\cref{fig:rq4_1} depicts a social network among the most productive authors for the analyzed literature.
The network nodes represent authors: the larger the node, the more papers the author has published.
The network edges represent collaborations among authors; the edge's thickness represents the number of collaborations between two authors.

Through the network and the table, it is possible to pinpoint the authors that can be considered the most productive in the field.
Mario Piattini has published 13 papers, followed by Perez-Castillo Ricardo with ten papers, and Shaukat Ali, Frank Leymann, and Tao Yue with nine papers each.

The most substantial collaboration cluster has been identified as the one formed by Mario Piattini and his collaborators Ricardo Perez-Castillo, Luis Jimenez-Navajas, Guido Peterssen, and Manuel Serrano.
These authors have a strong collaboration history and the highest number of published papers.

The second strongest collaboration cluster comprises Shaukat Ali, Paolo Arcaini, Tao Yue, and Xinyi Wang.
Although it is evident that the cluster size is smaller than other clusters, the strength of the collaborations is higher than any other identified cluster.

Frank Leymann also stands out as having many collaborations, co-authoring papers with Daniel Vietz, Joanna Barzen, Benjamin Weder, Manuela Weigold, and Marie Salm.
These authors form a solid cluster regarding published papers and the strength of the collaborations.

Another collaboration cluster comprises Enrique Moguel, Javier Rojo, David Valencia, Jose Garcia-Alonso, Javier Berrocal, and Juan Manuel Murillo.
However, this cluster shows a less strong relationship between the authors and fewer papers.
Another smaller but noticeable cluster comprises Jianjun Zhao, Zhongtao Miao, Pengzhan Zhao, and Shuhan Lan.

In addition to the strong collaboration clusters identified, it is worth noting that there are many tiny other groups and other one-time collaborations among the authors in the data, which may not be visible in the provided figures, but they likely occur between authors who have published occasionally together.

\begin{figure*}[!t]
        \centering
        \includegraphics[width=.7\columnwidth, height=!]{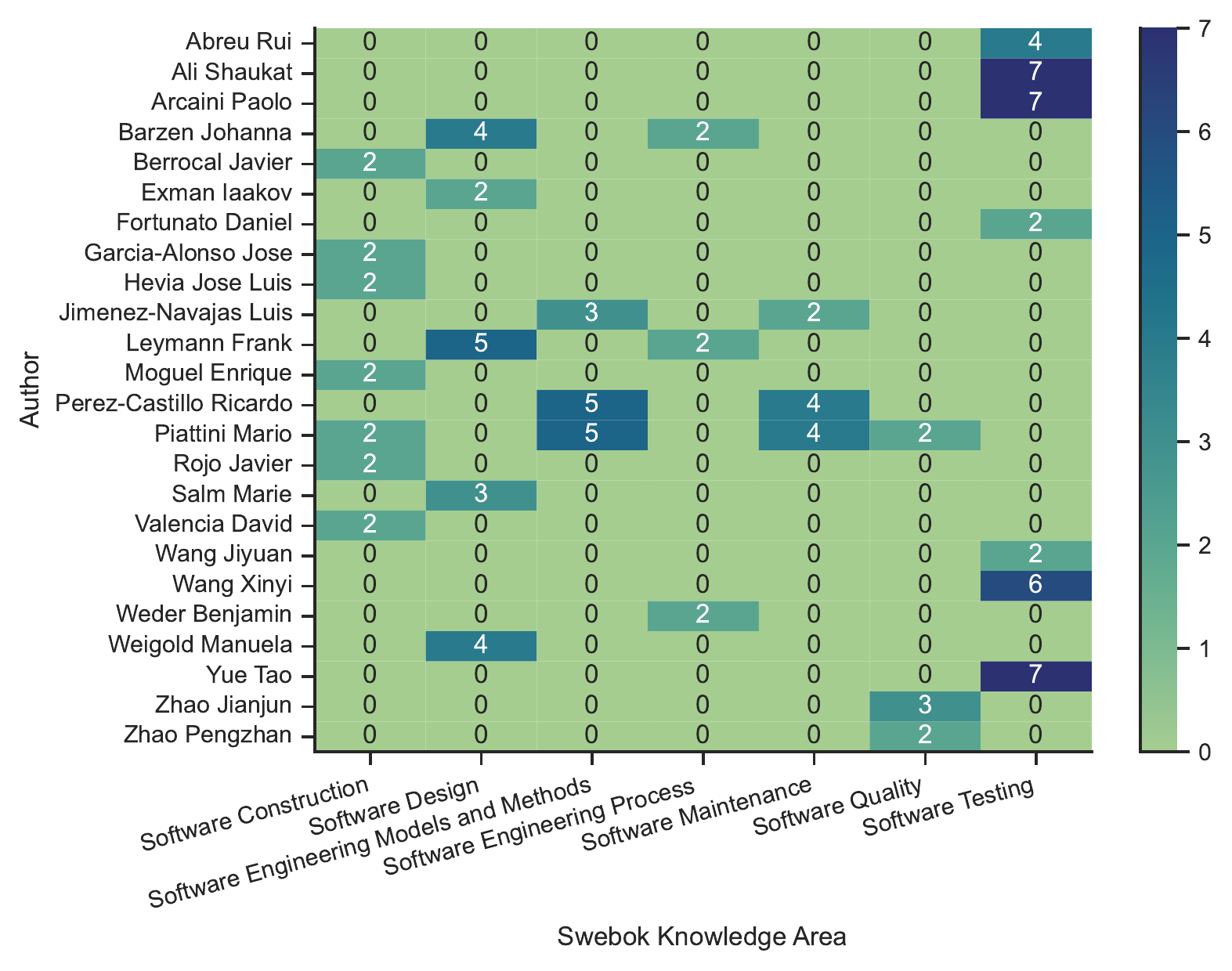}
        \caption{Top authors for each considered Knowledge Area. Mario Piattini is the leading author with the highest number of publications (13), covering many knowledge areas. Four authors published in multiple areas, while most focused on one.} \label{fig:rq4_2} 
\end{figure*}

\Cref{fig:rq4_2} presents a ranking of top authors in the field of quantum software engineering based on the number of publications related to the considered knowledge areas of the SWEBOK.

The top author in the table, with the highest number of publications overall, is Mario Piattini, as previously shown, who has published a total of 13 papers in the field. 
Piattini has contributed to a wide range of knowledge areas, with the most publications in Software Engineering Models and Methods (five papers) and Software Quality (2 papers). 
Other notable authors in the table include Xinyi Wang, with six publications in the field, primarily in the Software Engineering Process knowledge area, and Rui Abreu, who has published four papers in the Software Testing knowledge area.

Interestingly, almost all the authors have publications focused on a restricted number of knowledge areas. 
For example, Frank Leymann has published five papers in the Software Design knowledge area, suggesting a strong focus on this area of quantum software engineering, although contributing to the Software Engineering Process area as well. 
In contrast, authors such as Shaukat Ali and Jianjun Zhao have not published in any knowledge areas except for Software Testing and Software Quality, respectively.
Mario Piattini is the sole author who has published in more than two focus areas: software construction, software design, software engineering models and methods, and software quality.
Of the 22 authors listed, only four published papers in two or more focus areas. Specifically, Johanna Barzen published in Software Design and Software Engineering Process, Luis Jimenez-Navajas published in Software Engineering Models and Methods and Software Maintenance, Frank Leymann published in Software Design and Software Engineering Process, and Ricardo Perez-Castillo in Software Design and Software Maintenance.

\begin{summarybox}[RQ$_{4}$]
    The researchers most interested in Quantum Software Engineering (QSE) are Mario Piattini, Ricardo Perez-Castillo, Shaukat Ali, Frank Leymann, and Tao Yue, with Piattini being the most productive author in the field, with 13 publications. 
    The researchers are interconnected through collaborations, and several strong collaboration clusters have been identified, with the strongest being Piattini and his collaborators, Perez-Castillo, Jimenez-Navajas, Peterssen, and Serrano (\textbf{RQ$_{4.1}$}). 
    Most researchers have focused on a specific knowledge area regarding the distribution of researchers across different SE topics. 
    Piattini is the only author to have published in more than two areas (\textbf{RQ$_{4.2}$}).
\end{summarybox}

\subsection{RQ5: Venues}

\begin{figure*}[!ht]
        \centering
        \includegraphics[width=.7\columnwidth, height=!]{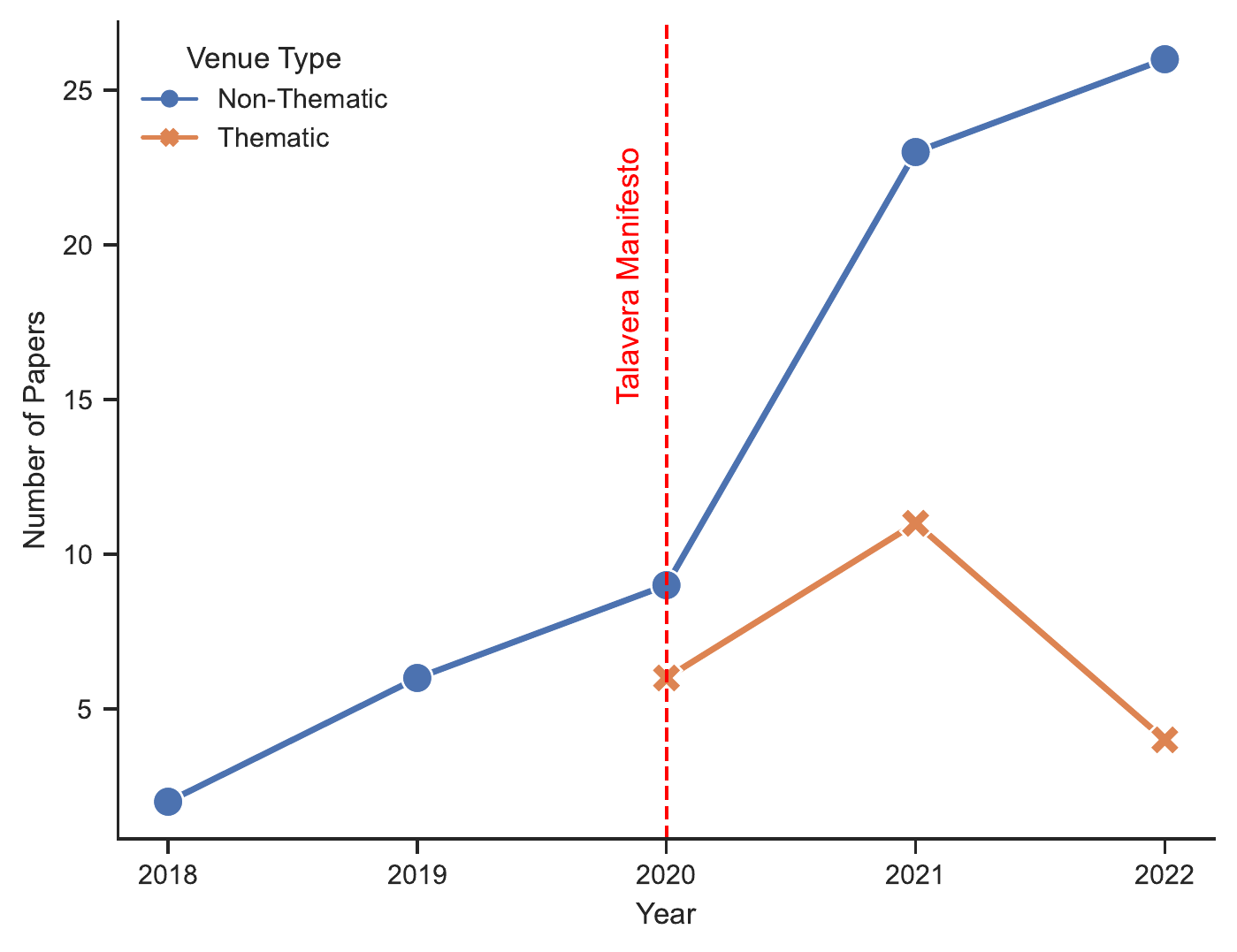}
        \caption{Number of papers published in both thematic and non-thematic venues over the years.} \label{fig:rq5_1} 
\end{figure*}

\begin{figure*}[!ht]
    \centering
    \includegraphics[width=.8\columnwidth, height=!]{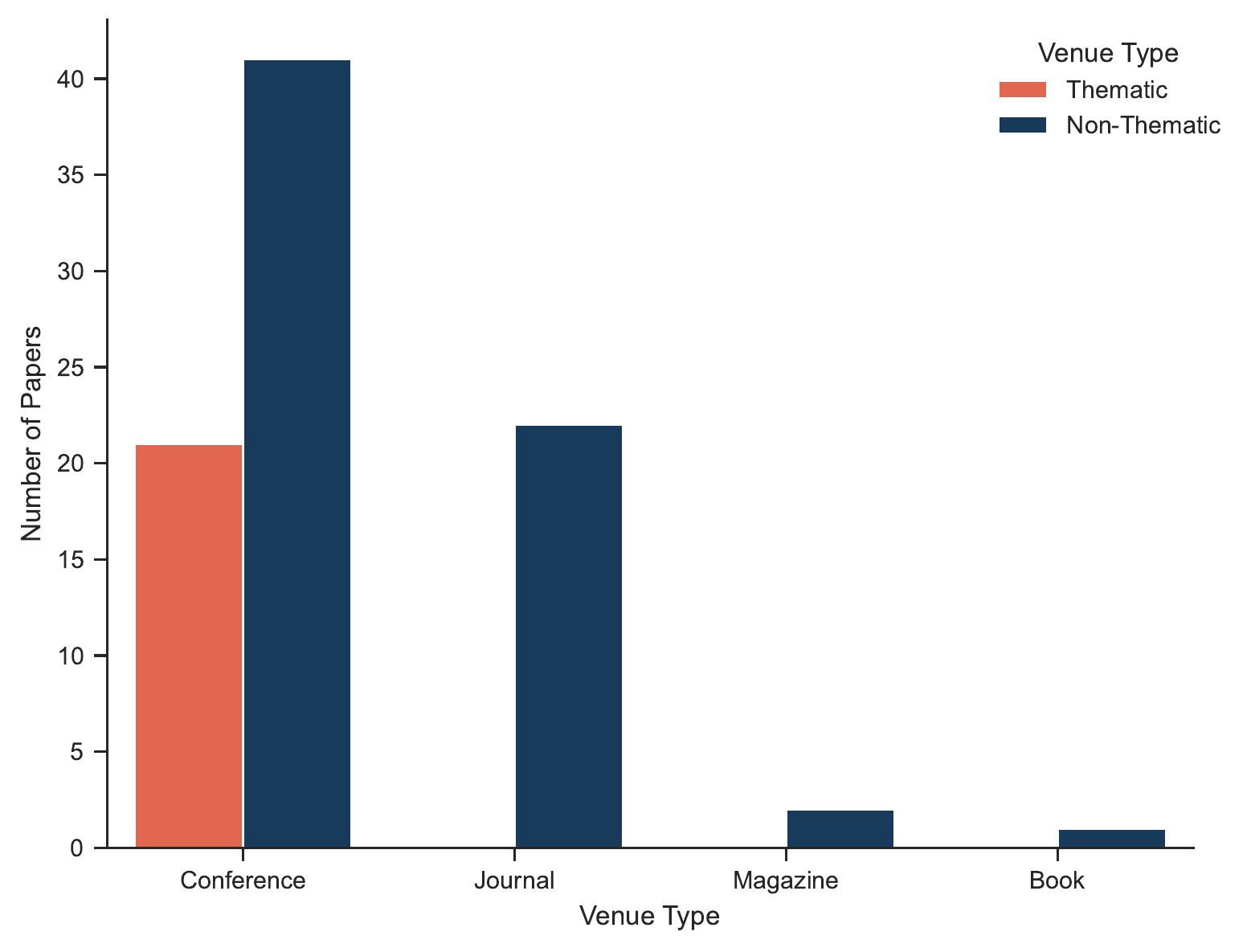}
    \caption{Papers by venue type.} \label{fig:rq5_2}
\end{figure*}

\Cref{fig:rq5_1} depicts the number of papers published in thematic and non-thematic venues over the years, while \Cref{fig:rq5_2} depicts the total number of publications by venue type.
\Cref{tab:venues} shows the top venues by number of papers.
Several insights can be made.
First, the field is active and diverse, with research published in various venues. 
The data shows that papers have been published in conferences (41 papers), journals (23 papers), magazines (two papers), and books (one paper).
This result suggests that the field is well-established and has a strong presence in multiple publication outlets.

\begin{table}[!ht]
\centering
\caption{Top venues by the number of papers. The venues are divided into thematic (\ie specialized in QSE) and non-thematic workshops. All thematic venues are reported, whilst non-thematic venues with more than 2 papers are reported.}
\label{tab:venues}
\begin{tabular}{p{0.5\columnwidth}rrp{0.2\columnwidth}}
\toprule
                                                                                            Venue & Thematic & Articles &                                                                                                                                                     References \\
\midrule
                                    International Workshop on Quantum Software Engineering (Q-SE) &      Yes &        9 & \citeR{Pontolillo2022-ep, Abreu2022-mm, Exman2021-yg, Perez-Castillo2021-jz, Campos2021-ri, Zhao2021-gx, Gemeinhardt2021-kt, N_Oldfield2022-lo, P_Zhao2021-iq} \\
                                      Quantum Software Engineering and Technology Workshop (QSET) &      Yes &        7 &                           \citeR{Scheerer2021-rw, Mykhailova2021-ss, Verduro2021-nk, Hevia_Oliver2020-le, Jimenez-Navajas2020-ew, Gomes2020-mw, Moguel2020-yc} \\
                                                                           ICSE Workshops (ICSEW) &       No &        4 &                                                                                   \citeR{Perez-Delgado2020-ed, Kruger2020-rc, Honarvar2020-gj, Barbosa2020-we} \\
                                                             Automated Software Engineering (ASE) &       No &        4 &                                                                                        \citeR{E_Mendiluze2021-ls, X_Wang2021-ew, P_Zhao2021-ou, J_Wang2021-jl} \\
                                                                                          Quantum &       No &        3 &                                                                                                           \citeR{LaRose2022-wt, LaRose2019-ti, Steiger2018-kg} \\
The International Conference on the Quality of Information and Communications Technology (QUATIC) &      Yes &        3 &                                                                                      \citeR{Jimenez-Navajas2020-jm, Jimenez-Navajas2021-eb, Cruz-Lemus2021-it} \\
             International Workshop on the Quantum Software Engineering and Programming (QANSWER) &      Yes &        1 &                                                                                                                                       \citeR{Peterssen2020-yr} \\
                                                 International Workshop on Quantum Software (QSW) &      Yes &        1 &                                                                                                                                \citeR{R_Perez-Castillo2022-ms} \\
\bottomrule
\end{tabular}
\end{table}

Conferences are the most preferred venue type and the only type showing thematic venues.
Among conference venues, ICSEW and ASE have published most of the papers in this field, with four papers each.
In general, many venues comprised papers about quantum software engineering, indicating that the field is active and diverse, with research published in many different outlets.

Journals have also published many papers in this field, with 23 papers. 
Quantum Journal, Journal of Systems and Software, Advances in Engineering Software, APEQS, Software Quality Journal, and SummerSOC are among the venues that have published two papers in this field. 
Several venues have published only one paper in this field, including top-tier journals and conferences, such as IEEE Transactions on Software Engineering, ICSME, and SANER.

The remaining non-thematic venues are represented by Magazines and books, which have published a relatively small number of papers in this field, \ie two and one respectively.

Thematic workshops are essential for publishing new ideas in the field.
Among all papers, 21 were published in thematic workshop conferences, representing a significant portion of the published papers. 
Among these, Q-SE and QSET have published the most papers, with nine and seven papers, respectively.

\begin{summarybox}[RQ$_5$]
    Since thematic venues represent a significant proportion of the publication venues, articles in QSE have been published in many other non-thematic venues, with a preference for conferences over others.
    The preferred non-thematic publication venues are ICSE Workshops and ASE, both having four publications each (\textbf{RQ$_{5.1}$}).
\end{summarybox}
\section{Discussion and Insights for Future Research} \label{sec:discussion}

\subsection{On the current research trends in QSE}

The resource distribution analysis reveals several key insights. 
First, the high prevalence of Solution Proposal papers (33 resources across all focus areas) suggests a strong demand for practical solutions. 
This result implies that researchers are actively working to address the challenges practitioners face and develop innovative approaches that can help improve software engineering practices.
The most challenging aspect is indeed related to testing and debugging quantum programs, for the intrinsic issues of the practice \cite{garcia2021quantum}.

Second, the few resources focusing on the software engineering processes (only three resources) highlight the need for further research.
It suggests that there is still much to be discovered regarding best practices, processes, and methodologies for software engineering. 
More research is needed to advance our understanding of this area.
Moreover, we saw that some knowledge areas in software engineering, such as software requirements, configuration management, and software engineering management, have been neglected and have not been the subject of any published papers. 

In particular, the lack of software project management can have significant implications for the socio-technical aspects of quantum software development, such as communication and collaboration among developers, which we can assume come from different backgrounds \cite{zhao2020quantum, piattini2021toward, ali2022software}. 
Effective management is critical to ensure that resources are effectively utilized, goals are aligned, and stakeholders are effectively engaged.

Such areas could have been neglected because they are perceived as less technically challenging or trending.
However, the issue could also be a prioritization choice of the researchers, who focus on solving more tangible and concrete problems, such as testing and debugging \cite{garcia2021quantum}.
Nevertheless, more research is necessary to address the issues and ensure all aspects of software engineering are adequately covered.

Third, Validation research resources are distributed relatively evenly across focus areas, with two to three resources each, highlighting its significance in software engineering. 
This type of research is important for providing evidence and supporting effective practices and solutions.
At the time of writing, however, empirical research in QSE is limited compared to other activities like solution proposals. 
Additionally, the discipline almost completely lacks industrial empirical studies.

Finally, the unequal distribution of research types in different focus areas (e.g., a higher concentration of Solution Proposals in Software Testing) underscores the importance of considering each area's unique characteristics and needs when planning research activities to help ensure that research efforts are directed toward the areas that will have the most significant impact.

\keyfindingsone{Several areas require further research. 
These include software engineering processes, software project management, neglected knowledge areas such as software requirements and configuration management, and the need for more academic and industrial empirical studies in quantum software engineering. 
Additionally, it is important to consider the unique characteristics and needs of each area when planning research activities to ensure that efforts are directed toward those that will have the most significant impact.}

\subsection{On the achieved results and studied technologies}

In \Cref{sec:results}, we presented the main results achieved by the analyzed resources and the main quantum technologies interested.
The results provide insights into the state of research in quantum software engineering and highlight the diversity of research in this field.

The technique was the most commonly achieved result type, with 22 resources reporting it.
As a newly born discipline, the field of quantum software engineering is still in its early stages of development. 
Thus, it is unsurprising that techniques are the most prominent result types reported in the resources analyzed.
In the early stages of a field's development, researchers tend to focus on developing methods and frameworks that can be used to tackle specific challenges because the field is still in its formative years, and it is essential to establish a solid foundation of techniques and tools that can be used to advance the field further. In quantum software engineering, the development of new techniques has been driven by the need to address various challenges in quantum software development \cite{zhao2020quantum, piattini2021toward, piattini2020quantum}.

The finding that a considerable number of resources focus on multiple or no specific quantum technologies can be attributed to the interdisciplinary nature of the field and the need to consider multiple technologies and perspectives. 
As said, QSE is a relatively new field, and researchers and practitioners are still exploring the potential of different quantum technologies and their integration with classical computing systems \cite{piattini2020talavera, zhao2020quantum}. 
The use of multiple or no specific technologies may aim to develop technology-agnostic methods that can be used with various quantum technologies, as advocated by the Talavera Manifesto \cite{piattini2020talavera}.

The resources that propose tools as the primary result in quantum software engineering are more focused on specific technologies because there is a practical need for concrete implementations in this field.
As a newly-born discipline, there is still much room for exploration and experimentation. 
Tools provide a tangible means for researchers and practitioners to test and validate their ideas and for users to put the theories into practice.
The concentration of resources focused on specific technologies like Qiskit indicates that this technology has become popular among the community, possibly due to its ease of use, user-friendly interface, and compatibility with various quantum computing hardware. 
This popularity will likely drive the development of many tools and techniques, further increasing its popularity and usability.
In addition to tools addressing developers' needs, tools represent a fundamental requirement for conducting experiments on codebases, which explains the significant effort devoted to developing tools early in a newly-born discipline.
In conclusion, the prominence of the tool result type and the concentration of resources focused on specific technologies highlights the practical and tangible nature of quantum software engineering and the importance of concrete implementations for advancing the field.

The low representation of result types such as Datasets, Guidelines, and Catalogs in the analyzed resources could be due to several reasons.
Firstly, the field of quantum software engineering is relatively new and still in its early stages of development. As a result, there might not be a strong emphasis on building extensive datasets, guidelines, or catalogs to support quantum software development. 
Instead, the focus is on developing techniques and tools necessary for software engineering practices in the quantum domain.
Additionally, collecting and preparing large datasets, guidelines, or catalogs in the quantum software engineering domain could be challenging because quantum technologies are still being developed and standardized. The associated data and information are constantly evolving, making it challenging to build robust and comprehensive datasets, guidelines, or catalogs that can be widely adopted and used by the research community.
Finally, the low representation of Datasets, Guidelines, and Catalogs could also be attributed to the current state of research in quantum software engineering. Many researchers in this field might focus more on exploring new techniques and tools rather than investing time and resources into building datasets, guidelines, and catalogs. 
As the field of quantum software development grows, it's crucial to have comprehensive and current resources to support it. 
However, the lack of such resources shows a significant research gap in quantum software engineering. As the discipline matures, the need for these resources is likely to become more evident, and more researchers may push in this direction.

\keyfindingsone{Further research in quantum software engineering should focus on filling the research gap highlighted by the low representation of result types such as Datasets, Guidelines, and Catalogs in the analyzed resources. 
Building comprehensive and up-to-date resources to support quantum software development could be essential in advancing the field further. Additionally, exploring new techniques and tools to tackle specific challenges and investigating the potential of different quantum technologies and their integration with classical computing systems should remain a priority for researchers and practitioners.}

\subsection{On the evolution of QSE}
The field of quantum software engineering has experienced rapid growth in recent years, as evidenced by the significant increase in publications in the discipline. 
The establishment of the discipline, through the publication of the Talavera Manifesto in 2020, has provided a clear definition and set of guidelines for future research and development in this field. 
This milestone has been crucial in recognizing quantum software engineering as a distinct discipline and will continue to play an essential role in shaping its future.

Early publications in quantum software engineering indicate that researchers and practitioners have already explored and developed the concept of quantum software engineering, even before its formal recognition. This result highlights the growing importance and recognition of the field and its potential to impact the future significantly.
The rapid increase in the number of publications, doubling from 2020 to 2021, suggests that there has been a significant increase in research and development in this field, and it is likely to continue to proliferate.
This increase in publications also suggests that quantum software engineering attracts more attention from researchers, practitioners, and investors, who recognize its potential to drive innovation and impact in various industries and domains.
While the data for 2022 only covers papers published until November of that year and may not be comprehensive, the significant increase in the number of publications offers valuable insights into the progress of quantum software engineering.

Based on the rapid growth and increasing interest in the field of quantum software engineering, it is likely that the discipline will continue to experience significant development in the near future. 
The growing recognition and importance of the field, the support and demand from various industries and domains, and the potential of quantum computing and quantum software to revolutionize the way we process and store data, optimize complex systems, and solve problems in various fields, will drive further investment and research in quantum software engineering.
This discipline is expected to have significant growth and development in the near future and will play a crucial role in shaping technology and innovation.

\keyfindingsone{The rapid growth and increasing interest in the field of quantum software engineering suggest that it will continue to experience significant development in the near future. 
To further advance the discipline, future research could explore the development of best practices, standards, and tools for quantum software engineering and investigate the potential applications of quantum software in various industries and domains.}

\subsection{On authors and collaborations in QSE}
In \Cref{sec:results}, we pointed out which are the most interested authors in QSE in the software engineering community, how they are related, and how they are distributed among the SWEBOK knowledge areas.

The social network analysis conducted on QSE authors has revealed several collaborative clusters with varying degrees of strength and cohesion.
These clusters suggest that QSE researchers tend to form closely-knit communities, which could hinder the exchange of knowledge and ideas between different groups.
To address this issue, initiatives to promote interdisciplinary collaboration among researchers in different collaboration clusters may be advantageous.
These initiatives may include hosting joint conferences or workshops to facilitate the exchange of knowledge and ideas, potentially leading to novel insights and discoveries in the field of QSE.
Such efforts may also aid in building more robust and diverse research communities, enabling the development of innovative approaches to QSE problems.

The analysis of the knowledge areas of QSE authors reveals that most researchers focus on specific areas rather than the entire spectrum of QSE knowledge. 
While some authors have published many papers, they mostly focus on a few knowledge areas, such as software design and engineering processes. 
In contrast, other areas, such as software testing and maintenance, are less well-represented, which suggests that there is a need for more research on these underrepresented areas to create a more holistic understanding of quantum software engineering.
In addition, very few authors publish in multiple areas suggesting that there may be a lack of interdisciplinary collaboration between quantum computing and software engineering experts. 
Encouraging more collaboration between these groups could help to create a more balanced research focus in quantum software engineering.

\keyfindingsone{The social network analysis of QSE authors highlights the need for initiatives to promote interdisciplinary collaboration between closely-knit research communities, which could lead to novel insights and discoveries in QSE and help build more robust and diverse research communities. 
Additionally, the analysis of QSE authors' coverage of SWEBOK knowledge areas reveals a need for more research on underrepresented areas such as software testing and maintenance.
Encouraging collaboration between quantum computing and software engineering experts could help create a more balanced research focus in quantum software engineering.}

\subsection{On the publication trends in QSE}
The field of quantum software engineering is a growing and dynamic field that is seeing an increase in the number of publications in various venues.
Despite the significant share represented by thematic venues, more recently, non-thematic venues were preferred.
The preference for non-thematic venues over thematic ones may suggest the field is leaving its niche discipline status and permeating into well-established software engineering publication outlets. 
This result can be seen as a positive trend, indicating that quantum software engineering is gaining recognition and attention from the broader software engineering community.
Reaching a broader audience through non-thematic venues may help to promote the field and increase its visibility, leading to more collaboration and exchange of ideas with other software engineering sub-fields. 
Furthermore, publishing in non-thematic venues may also increase the impact of the research, as it can reach a more significant number of readers who may not be familiar with the specific field.
Moreover, publishing in non-thematic venues also allows QSE researchers to showcase their work to a wider audience and demonstrate their research's relevance and potential impact in the broader context of software engineering. 
It may also help attract new researchers and provide a platform for interdisciplinary collaboration.
In conclusion, such evolution could give birth to new research directions and the advancement of existing ones and could further support the establishment of the field.

It is important to note that the data presented covers up to December 2022 (excluded) and may not accurately represent the trends in the field over the entire year. 
It could be beneficial to gather data on other factors that might influence the publication trends in this field, such as funding or resource availability.

\keyfindingsone{The increasing preference for non-thematic publication venues in quantum software engineering suggests that the field is gaining recognition and attention from the broader software engineering community, which could lead to new research directions and interdisciplinary collaborations. 
However, it would be beneficial to investigate other factors influencing publication trends, such as funding or resource availability. 
Furthermore, more research is needed to understand how the integration of quantum software engineering with the broader software engineering community can be leveraged further to support the growth and recognition of the field.}
\section{Conclusions} \label{sec:conclusion}
In this systematic mapping study, we aimed to provide a comprehensive overview of the current state of research in Quantum Software Engineering (QSE).
Through an extensive analysis of 87 studies, our main contributions are: (i) a comprehensive synthesis and analysis of the research conducted in the field of quantum software engineering, which might be useful to researchers and practitioners to learn how software engineering and quantum computing have been combined so far;
(ii) a systematic mapping of the venues and research groups that are currently focusing on quantum software engineering, which may be useful to newcomers and researchers interested in starting their path in this research field to discover relevant venues and potential collaborators;
(iii) A research roadmap that highlights the major achievements that further research should pursue;
(iv) An online appendix reporting all the material used to conduct our systematic mapping study, which researchers might use to build on top of our research and extend our findings\footnote{Appendix available at \url{https://figshare.com/articles/online_resource/The_Quantum_Frontier_of_Software_Engineering_A_Systematic_Mapping_Study/22263448}}.

Our results indicate that QSE research is still in its early stages, focusing on software testing and neglecting some knowledge areas, such as software engineering management. 
The most reported results types are techniques, empirical, and positions, with Qiskit being the most commonly studied technology. 
We also observed a growing interest in QSE within the research community, with a speedy increase in published papers between 2020 and 2021.
Regarding researchers, we identified the most productive authors, the main collaboration clusters, and the distribution of researchers across different SE topics. This information can help to identify potential collaborators and promote further research in QSE.
Finally, our study highlights the need for more empirical studies and a better distribution of research efforts across different SE topics. We also encourage more non-thematic publication venues to consider QSE papers to broaden the research community's knowledge and reach.
Our study provides valuable insights into the development and evolution of the research community, contributing to the advancement and growth of QSE.

This systematic mapping study highlights potential future directions for research in Quantum Software Engineering (QSE), especially in the neglected areas of software engineering management practices and quantum software maintenance. 
Future research could focus on developing effective strategies and tools for managing the software development process and maintaining reliable and performing quantum software over time. 
Additionally, there is a need for more empirical studies and the development of appropriate metrics to provide a more rigorous empirical basis for future research in QSE.
Future research could investigate effective strategies for managing the software development process in the context of quantum computing to address the neglect of software engineering management practices in QSE. 
Such efforts could involve exploring the unique challenges and opportunities of quantum software engineering, identifying effective strategies for managing the development process, and evaluating the effectiveness of different software engineering practices and tools.
Research on quantum software maintenance could also provide new insights and discoveries. 
Future studies could explore the challenges and opportunities of maintaining quantum software over time, including issues such as version control, bug fixes, and updates. 
Developing appropriate metrics could also help provide a more rigorous empirical basis for future research in QSE, allowing for the evaluation of code quality, performance, and reliability.

\section*{Appendix}
Our online appendix is avalilable at the following link \url{https://figshare.com/articles/online_resource/The_Quantum_Frontier_of_Software_Engineering_A_Systematic_Mapping_Study/22263448}.

\section*{Acknowledgement}
This work has been partially supported by the EMELIOT national research project, funded by the MUR under the PRIN 2020 program (Contract 2020W3A5FY). Fabio gratefully acknowledges the support of the Swiss National Science Foundation through SNF Projects No. PZ00P2\_186090.

\balance

\bibliographystyle{elsarticle-num-names}
\bibliography{biblio}

\bibliographystyleR{elsarticle-num-names}
\bibliographyR{resources}

\end{document}